\documentclass[12pt]{article}

\usepackage{color}
\usepackage{kotex}
\usepackage{graphicx}
\usepackage{epsfig}
\usepackage{bm}
\usepackage{amssymb, amsmath, amsthm}
\usepackage{verbatim}
\usepackage{natbib}
\usepackage{authblk}
\usepackage{wasysym}
\usepackage{latexsym}
\usepackage{amsfonts}
\usepackage{subcaption}
\usepackage{listings}
\usepackage{booktabs}
\usepackage{multirow}
\usepackage{booktabs}
\usepackage{bbold}
\usepackage{adjustbox}
\usepackage{rotating}
\usepackage{ulem}
\usepackage{xcolor}

\newcommand{\biblist}{\begin{list}{}
{\listparindent 0.0cm \leftmargin 0.50cm \itemindent -0.50 cm
\labelwidth 0 cm \labelsep 0.50 cm
\usecounter{list}}\clubpenalty4000\widowpenalty4000}
\newcommand{\ebiblist}{\end{list}}

\newtheorem{exmp}{Example}[section]
\newtheorem{defn}{Definition}[section]
\newtheorem{prop}{Proposition}[section]
\newtheorem{thm}{Theorem}[section]
\newtheorem{cor}{Corollary}[section]

\newcommand{\E}{\operatorname{E}}
\newcommand{\V}{\operatorname{var}}
\newcommand{\T}{\operatorname{T}}
\newcommand{\B}{\boldsymbol}

\newcommand{\OP}{\operatorname}
\newcommand{\bx}{\boldsymbol{x}}

\title{Maximum Likelihood Imputation} 
\author{Jeongseop Han \and Youngjo Lee \and Jae Kwang Kim}
\date{} 
\begin{document}

\baselineskip .3in
\maketitle 

\begin{abstract}
Maximum likelihood (ML) estimation is widely used in statistics. The h-likelihood has been proposed as an extension of Fisher's likelihood to statistical models including unobserved latent variables of recent interest. Its advantage is that the joint maximization gives ML estimators (MLEs) of both fixed and random parameters with their standard error estimates. However, the current h-likelihood approach does not allow MLEs of variance components as Henderson’s joint likelihood does not in linear mixed models. In this paper, we show how to form the h-likelihood in order to facilitate joint maximization for MLEs of whole parameters. We also show the role of the Jacobian term which allows MLEs in the presence of unobserved latent variables. To obtain MLEs for fixed parameters, intractable integration is not necessary. As an illustration, we show one-shot ML imputation for missing data by treating them as realized but unobserved random parameters. We show that the h-likelihood bypasses the expectation step in the expectation-maximization (EM) algorithm and allows single ML imputation instead of multiple imputations. We also discuss the difference in predictions in random effects and missing data.
\end{abstract}

\baselineskip .3in

\baselineskip .3in

\baselineskip .3in

\baselineskip .3in

\baselineskip .3in

\baselineskip .3in

\baselineskip .3in

\baselineskip .3in

\newpage

\section{Introduction}
\label{section:ML_intro}

Missing data are prevalent in statistical problems,  but ignoring them can lead to erroneous results \citep{little2019, kimshao2021}. Imputation is a
popular technique for dealing with missing data. However, if imputed data are treated as observed, the use of the standard statistical procedure could result in erroneous inference, giving a biased estimator
with an underestimated standard error estimator. Multiple imputation has
been proposed by \cite{rubin1987} to address the uncertainty associated with
imputation. However, it requires the self-consistency  conditions 
\citep{wang98,meng94,yang16}, which may not necessarily hold. An alternative method by \cite{kim11} is fractional imputation. 

ML estimation of \cite{fisher1922} is widely accepted in estimating fixed parameters. Missing data can be
viewed as unobserved random parameters \citep{leenelderpawitan2017} so that
imputation can be viewed as a prediction of random parameters, namely missing data. It necessitates an extension of the Fisher likelihood to statistical models that include unobserved random variables 
\citep{bergerwolpert1984, butler1986}. \cite{lee1996} intended an extension of ML estimation to models
with unobserved random parameters via h-likelihood, defined on a particular scale of random parameters in the linear predictor. However, they confronted severe objections due to difficulties as \cite{bayarri1988} showed that ML estimation of extended likelihood  often provides nonsensical estimation for both fixed and random parameters. Furthermore, \cite{firth2006} noted that the linear predictor to form the h-likelihood might not be necessarily well defined. All the counterexamples against the h-likelihood, for examples in 
\cite{little02}, are associated with a wrong choice of scale to form h-likelihood. \cite{little2019} described the current status of h-likelihood ``Unlike maximization of the marginal likelihood of \cite{fisher1922}, maximization of an extended likelihood does not generally give consistent estimates of the parameters \citep{breslow95} ... \cite{leenelder2001} and 
\cite{leenelderpawitan2006} propose maximizing a ``modification'' ... which is the correct ML approach. For more details, see \cite{leenelder2009} and the discussion, particularly \cite{meng2009}.'' The success of h-likelihood approach looks coincidental, so that \cite{meng2009} tried a rigorous theoretical justification for the use of h-likelihood by showing its Bartlett identities. But he ended up highlighting the  difficulty caused by the difference between fixed and random parameter estimations. Thus, the benefit of using
h-likelihood has not been well accepted yet. This paper establishes the original aim of the h-likelihood whose maximization without any modification provides correct ML estimation and ML imputation by giving rigorous justifications.

\cite{leenelderpawitan2006} defined h-likelihood precisely, but they have not fully exploited its usefulness. For example, an immediate drawback of the current h-likelihood is that it does not allow MLEs of variance components as Henderson's \citet{henderson1959} joint likelihood does not. So \cite{leenelderpawitan2017} use a modification to obtain MLEs of variance components, etc. We need to reformulate the h-likelihood in a thoroughly consistent way to avoid modification. Jacobian terms do not play any role in Fisher’s (1922) ML estimation of fixed parameters. However, in models with random parameters, as we shall show, Jacobian terms play a key role in ML estimation. This property has not been well known yet in literature. We clarify the role of the Jacobian term in defining h-likelihood. Currently, the h-likelihood has been defined mainly for random effect models, where linear predictors are defined \citep{lee1996}. To illustrate our proposal for a  much wider class of models, we consider the  imputation problem, which does not require a linear predictor, as noted by \cite{firth2006}, and encounters difficulties in ML estimation of random parameters, as noted by \cite{meng2009}. We clarify that the definitions of canonical scale and canonical function are
keys to leading valid ML estimation on both fixed and random parameters without any modification in h-likelihood. 

In Section \ref{section:ML_basic}, we describe the basic setup for missing data problem. In Section \ref{section:ML_hlik}, we define the h-likelihood by using canonical scale and canonical function in terms of Jacobian term. Moreover, properties of MLEs for fixed and random parameters by using the h-likelihood are examined. In Section \ref{section:ML_canonicalization}, we propose the weak canonical scale based on the Laplace approximation. The weak canonical scale can give proper ML imputation when the canonical scale is unknown. In Section \ref{section:ML_ML}, we propose the ML imputation by using the MLE for random parameters. Illustrative examples in Section \ref{section:ML_examples} show the usefulness of the h-likelihood in the missing data problem.

\section{Basic Setup} 
\label{section:ML_basic}
Assume that we have a study variable $Y$ with dominating measure $\mu$ and a covariate vector $\boldsymbol{X}$. The study variable $Y$ is subject to missingness and the covariates are always observed. Assume  further that there are $n$ independent and identically distributed realizations of $(\boldsymbol{X},Y,\delta )$, denoted by $\{ (\boldsymbol{x}_{i},y_{i},\delta
_{i}):i=1,\ldots ,n \} $, where  $\delta _{i}$ is the missingness indicator defined by $\delta
_{i}=1$ if $y_{i}$ is observed and $\delta_i=0$ otherwise. We are interested in estimating $\eta= \E(Y)$ from the observed data. 

Under existence of missing data, an imputation estimator of $\eta$ can be written as 
$$ \hat{\eta}_{\OP{I}} = \frac{1}{n} \sum_{i=1}^n \left\{ \delta_i y_i + (1-\delta_i) \hat{y}_i \right\}  $$
where $\hat{y}_i$ is the imputed value of $y_i$. To predict realized values $y_i$ of unobserved missing data, we consider a frequentist approach using the ML imputation. The current procedure for ML imputation can be described as follows: 
\begin{description} 
\item{Step 1:} Estimate $\psi$ by maximizing the observed likelihood 
\begin{equation}
L_{m}(\psi )=f_{\psi }\left( y_{\OP{obs}},\delta \mid \boldsymbol{x}\right) =\int f_{\psi}(y_{\OP{obs}},y_{\OP{mis}},\delta \mid \boldsymbol{x})dy_{\OP{mis}},  \label{eq:marg_lik}
\end{equation}
where 
$f_{\psi }(y_{\OP{obs}},y_{\OP{mis}},\delta | \boldsymbol{x})$ is the joint density function of $\left( y_{\OP{obs}},y_{\OP{mis}},\delta \right) $ given $\boldsymbol{x}$ with fixed unknown parameter $\psi $ and  $(y_{\OP{obs}}, y_{\OP{mis}})$ is the observed and missing part of the complete data $y_{\OP{com}}=(y_1, \ldots, y_n)$, respectively. 
\item{Step 2:} For each $i$ with $\delta_i=0$, obtain a predictor of $y_i$ 
\begin{equation} 
 \hat{y}_i = \int y f( y \mid \bx_i, \delta_i=0; \hat{\psi}) d \mu(y) = \E_{\hat{\psi}} \left ( Y_{i} \mid \B{x}_{i}, \delta_{i} = 0 \right ),
 \label{mean}
 \end{equation} 
where $\hat{\psi}$ is the MLE of ${\psi}$ obtained from Step 1. 
\end{description}
We use subscript $m$ in the observed likelihood in (\ref{eq:marg_lik}) to emphasize that the likelihood is developed from the marginal density  of the observed data. 
\cite{robins00} and 
\cite{kimshao2021} present some asymptotic properties of the imputation estimator under ML imputation.
 The above two-step imputation procedure, however, is computationally {involved} as the ML estimation of the fixed parameter $\psi$ is often based on the iterative procedure such as EM algorithm \citep{dempster77}.  
However, such a conditional mean imputation in (\ref{mean}) does not necessarily give the best prediction in terms of maximizing the predictive distribution. For example, if $y$ is categorical, the conditional mean is not necessarily categorical. 

In this paper, instead of using the conditional mean imputation in (\ref{mean}), we propose using conditional mode of the h-likelihood given by
\begin{equation*}
 \hat{y}_{\OP{mis}} = \mbox{arg} \max_{y_{\OP{mis}}}  H(\hat{\psi}, y_{\OP{mis}})
 \end{equation*} 
 in the next section. In many practical situations, the conditional mode imputation is attractive as it respects the ``maximum likelihood'' principle by treating the unobserved $y$ values  as realized random parameters. By treating $y_{\OP{mis}}$ as the random parameters and applying the usual ML procedure, we can obtain imputed values, namely ML imputation, that adhere to the frequentist principle to the greatest extent possible. An immediate practical advantage is that one-shot imputation directly allows the ML estimation of fixed parameters. For one-shot imputation to be meaningful, as we shall show, it estimates the canonical function to predict future (or missing) variable, which resolves summarizability problem raised by \cite{meng2009}.
 
Naively treating the missing observations as unknown parameters will be subject to biased estimation, which is well known as pointed out by \cite{neyman1948}. Thus, we employ a technique known as h-likelihood \citep{lee1996}, to circumvent this issue and obtain valid inferences.  \cite{yun2007} studied the h-likelihood approach to estimate fixed parameters in missing data problems. We introduce the ML imputation of missing data and conduct a more systematic investigation, elucidating the mysteries of h-likelihood in general. 

\section{H-likelihood}
\label{section:ML_hlik}

In this paper, we rearrange the indices as $\delta _{i}=1$ for $i=1,\ldots ,n_{\OP{obs}}$ and $=0$ for $i=n_{\OP{obs}}+1,\ldots ,n$ where $n_{\OP{obs}%
}=\sum_{i=1}^{n}\delta _{i}$, i.e., the first $n_{\OP{obs}}$ responses are observed and remaining $n_{\OP{mis}}=n-n_{\OP{obs}}$ responses are not observed. Missing data can be viewed as prediction of future data which are not observed yet. By treating $y_{\OP{mis}}$ as random parameters, the complete-data log-likelihood is an extended log-likelihood 
\begin{eqnarray*}
\ell _{e}(\psi ,y_{\OP{mis}}) &=&\log L_{e}(\psi ,y_{\OP{mis}})=\log
f_{\psi }(y_{\OP{obs}},y_{\OP{mis}},\delta \mid \boldsymbol{x}) \\ 
&=&\sum_{i=1}^{n_{\OP{obs}}}\log f_{\psi }(y_{i},\delta _{i}=1\mid %
\boldsymbol{x}_{i})+\sum_{i=n_{\OP{obs}}+1}^{n}\log f_{\psi }(y_{\OP{mis}%
,i},\delta _{i}=0\mid \boldsymbol{x}_{i}).   \notag 
\end{eqnarray*}
 Extended likelihood principle \citep{bjornstad1996}
states that $L_{e}(\psi ,y_{\OP{mis}})$ carries all the information in the data about unknown parameters $\psi $ and $y_{\OP{mis}}$. 

\cite{lee1996} proposed the h-likelihood for ML estimation on both fixed and random parameters. Due to a Jacobian term,
unlike a transformation of fixed parameter $\psi $, a nonlinear transformation of random parameter $v=g(y_{\OP{mis}})$ changes the extended likelihood 
\[
L_{e}\left( \psi ,v\right) =L_{e}\left( \psi ,y_{\OP{mis}}\right) \left| 
\frac{\partial y_{\OP{mis}}}{\partial v}\right| .
\]
Here if the joint maximization of $L_{e}\left( \psi ,v\right) $ gives the MLE of $\psi $, that of $L_{e}\left( \psi ,y_{\OP{mis}}\right) $ cannot give the MLE of $\psi$. It means that specifying the scale of a random parameter in defining the h-likelihood is important to obtain MLEs via its maximization. In this paper, we elaborate on how to use the Jacobian term to form such an h-likelihood. 

Following \cite{leenelderpawitan2017},  the predictive likelihood of random
parameter $v$ can be defined as 
\[
L_{p}\left( v \mid \mathcal{D};\psi \right) \equiv f_{\psi }\left( v\mid \mathcal{D},%
\boldsymbol{x}\right) =f_{\psi }\left( v,\mathcal{D}\mid \boldsymbol{x}%
\right) /f_{\psi }\left( \mathcal{D}\mid \boldsymbol{x}\right) ,
\]
where  $\mathcal{D}=\{y_{\OP{obs}},\delta \}$ and subscript $p$ is used to emphasize  the predictive likelihood for $v$.   Thus, the marginal likelihood is expressed as 
\begin{equation}
  L_{m}(\psi)= \frac{ L_{e}(\psi ,v)}{L_{p}\left( v \mid \mathcal{D};\psi \right)}. 
  \label{eq6}
\end{equation}
Given $\psi$, let 
\begin{equation}
    \tilde{v} = \tilde{v}(\psi, \mathcal{D}, \B{x})=\OP{arg}\OP{max}_{v}L_{e}(\psi,v)=\OP{arg}\OP{max}_{v}L_{p}\left( v \mid \mathcal{D};\psi\right) 
\label{vtilde} 
\end{equation}
be the common mode of the extended likelihood and the predictive likelihood. Note that the common mode $\tilde{v}(\psi, \mathcal{D}, \B{x})$ is a function of both parameter and data. However, we denote it as $\tilde{v}$ for notational convenience. Evaluating the marginal likelihood in (\ref{eq6}) at  $v=\tilde{v}$ leads to 
\begin{equation}
L_{m}(\psi )= \frac{L_{e} ( \psi ,\tilde{v} ) }{L_{p} ( \tilde{v} \mid \mathcal{D};\psi ) }.  \label{eq:relationship_lik}
\end{equation}
If both $L_{e}(\psi, v)$ and $L_{p}(v|\mathcal{D};\psi)$ are explicitly available, at least at the mode $\tilde{v}$, the MLE for $\psi$ is immediately obtained from (\ref{eq:relationship_lik}). However, both are not often available. 

\begin{defn}
\label{defn:canonical_function} If a scale $v=g(y_{\OP{mis}})$ satisfies 
\begin{equation}
L_{e} ( \psi ,\tilde{v} )\propto L_{m}(\psi ), 
\label{eq:def1}
\end{equation}
the $v$-scale is called the \textit{canonical scale} and the mode $\tilde{v}$ is called the \textit{canonical function}. The extended likelihood defined on the canonical scale $v$ is called the \textit{h-likelihood}, 
\begin{equation*}
H(\psi ,v)=L_{e}(\psi ,v).
\end{equation*}
By combining (\ref{eq:relationship_lik}) and (\ref{eq:def1}), $L_{p} ( \tilde{v} | \mathcal{D};\psi )$ 
does not depend on $\psi$ if $v$-scale is canonical, i.e. information neutral with respect to $\psi$ at the mode $\tilde{v}$.
\end{defn}
Here, we emphasize defining the h-likelihood with different parametrization of a random parameter. Let $\hat{\zeta}$ be the MLE of $\zeta =k(\psi )$ under the transformation $k(\cdot )$. Then, the MLE $\hat{\psi}=k^{-1}(\hat{\zeta})$ is invariant with respect to the transformation. Similarly, the MLE of a parameter from the h-likelihood is transformation invariant. That is, we can treat $v$ as if it is a fixed parameter after defining the h-likelihood in the sense that
\begin{equation}
\label{eq:hlik_invariant}
    H(\psi, y_{\OP{mis}}) = H \{ \psi, g^{-1} (y_{\OP{mis}}) \} = H(\psi, v)
\end{equation}
\citep{leenelder2005}. Here, we denote $H(\psi, y_{\OP{mis}})$ the h-likelihood in terms of $y_{\OP{mis}}$ as (\ref{eq:hlik_invariant}), whereas $L_{e}(\psi, y_{\OP{mis}})$ indicates the extended likelihood in which the canonical scale is yet unknown. From (\ref{eq:hlik_invariant}), the conditional mode of $y_{mis}$ is defined by
\begin{equation}
    \tilde{y}_{\OP{mis}} = \OP{arg} \max_{y_{\OP{mis}}} H(\psi, y_{\OP{mis}}) = g^{-1} (\tilde{v}).
\label{eqn9} 
\end{equation}
If the transformation $g(\cdot)$ is not linear, we get  $\tilde{y}_{\OP{mis}} \neq \OP{arg} \max_{y_{\OP{mis}}} L_{e}(\psi, y_{\OP{mis}})$. Thus, under the canonical  condition (\ref{eq:def1}), MLEs of both fixed and random parameters can be obtained by maximizing $H(\psi, v) = L_{e} ( \psi ,v )$. 

\cite{leenelderpawitan2017} gave a correct definition of canonical scale above, but have not exploited it to form the h-likelihood. We  now state a sufficient condition for the canonical property in (\ref{eq:def1}) as follows.  

\begin{prop}
\label{prop:alt_def_canonical}
If a transformation $v=g(y_{\OP{mis}})$ with bijective, differentiable function 
$g(\cdot )$ satisfies 
\begin{equation*}
\left| \frac{\partial v}{\partial y_{\OP{mis}}}\right| _{v=\tilde{v}%
}\propto L_{p}(\tilde{y}_{\OP{mis}}\mid \mathcal{D};\psi ),
\end{equation*}
where $\tilde{y}_{\OP{mis}}=g^{-1}(\tilde{v})$ and $\tilde{v}$ is defined in (\ref{vtilde}),  the canonical property in (\ref{eq:def1}) is satisfied. 
\end{prop}
Proposition \ref{prop:alt_def_canonical} gives further interpretation about Definition \ref{defn:canonical_function}.
\begin{equation}
L_{m}(\psi )=\frac{L_{e}(\psi ,\tilde{y}_{\OP{mis}})}{L_{p}(\tilde{y}_{\OP{mis}}\mid \mathcal{D};\psi )}\propto L_{e}(\psi ,\tilde{y}_{\OP{mis}})\left| \frac{\partial y_{\OP{mis}}}{\partial v}\right| _{v=\tilde{v}}=L_{e}\left( \psi ,\tilde{v}\right) = H(\psi, \tilde{v}).  \label{eq:Jacobian_canonical}
\end{equation}
Moreover, it shows how the canonical scale allows ML estimation. Now, we first study the {ML} estimation of the fixed parameter using h-likelihood.  

\subsection{MLE of Fixed Parameter}

Equation (\ref{eq:Jacobian_canonical}) characterizes the canonical scale which allows the ML estimation. 
\begin{thm}
\label{thm:canonical_existence} Suppose that the predictive likelihood $%
L_{p}(y_{\OP{mis}} | \mathcal{D};\psi )$ is unimodal with respect to $y_{%
\OP{mis}}$. Then, there exists the canonical scale to form the
h-likelihood.
\end{thm}
Theorem \ref{thm:canonical_existence} states a sufficient condition for the  existence of a canonical scale. When an explicit form of the canonical scale is not available, we present a way of defining a weak canonical scale based on the Laplace approximation in Section \ref{section:ML_canonicalization}. For now, we assume that an explicit form of the canonical scale $v=g(y_{\OP{mis}})$ is known. The following theorem shows how to obtain the MLE of fixed parameter and also its variance estimator.

\begin{thm}
\label{thm:hessian_identity} (i) The MLE of $\psi$ can be obtained by
solving the score equation 
\begin{equation*}
\frac{\partial \ell_{m} }{\partial \psi }=\frac{\partial }{\partial \psi }h\left( \psi ,\tilde{v} \right ) =\frac{%
\partial h}{\partial \psi }\Big |_{v=\tilde{v}}=0,  
\end{equation*}
where $h=\log H(\psi ,v)$ and $\ell_{m} =\ell_{m} (\psi )=\log L_{m}(\psi )$. 

(ii) The variance estimator of the MLE can be obtained from the Hessian matrix
of the h-likelihood as 
\begin{equation*}
\hat{I}^{\psi \psi }=I^{\psi \psi }\big |_{\psi =\hat{\psi}},~I^{\psi \psi
}=\left( -\frac{\partial ^{2}\ell_{m} }{\partial \psi \partial \psi ^{T}}\right)
^{-1},  
\end{equation*}
where the definition of $I^{\psi \psi }$ is in Appendix.
\end{thm}
To compare the h-likelihood approach with the EM algorithm, note that  
\begin{equation*}
\frac{\partial \ell_{m}(\psi)}{\partial \psi} = \E_{\psi} \left \{ \frac{\partial}{\partial \psi}  \ell_{e}(\psi, y_{\OP{mis}})  \Big | \mathcal{D}, \B{x}  \right \}.
\end{equation*}
This equality is called the mean score theorem \citep{louis82}. The EM algorithm \citep{dempster77}
obtains the solution to $\partial \ell_{m}(\psi)/\partial \psi =0$ by  
\begin{equation} 
\psi^{(t+1)} \leftarrow \mbox{solve } 
\E_{\psi^{(t)}} \left \{ \frac{\partial}{\partial \psi}  \ell_{e}(\psi, y_{\OP{mis}}) \Big | \mathcal{D}, \B{x} \right \}=0.\label{eq:em_score}
\end{equation} 
The h-likelihood approach gives the MLE of the fixed parameter without requiring the E-step in (\ref{eq:em_score}) which is often computationally intensive.

\subsection{MLE of Random Parameter}
\label{section:ML_MLE_random}
If we let $y_{\OP{mis}}$ be the unobserved part of the data, the missing data problem becomes a prediction problem. To understand Meng's point in \cite{meng2009}, assume that $y_{\OP{obs}}$ and $y_{\OP{mis}}$ are independent and the scale $v=g(y_{\OP{mis}})$ is the canonical scale. {Prediction of future data can be viewed as missing data problem where $y_{t+1}, \ldots, y_{t + n_{\OP{mis}}}$ are future data at the present time $t=n_{\OP{obs}}$.} \cite{meng2009} showed that 
\[
\hat{v}-v=g\left( \hat{y}_{\OP{mis}}\right) -g\left( y_{\OP{mis}}\right)
=g^{\prime }\left( \tilde{y}_{\OP{mis}}\right) \left( \hat{y}_{\OP{mis}%
}-y_{\OP{mis}}\right) +R_{n_{\OP{obs}}},
\]
where 
\begin{equation*}
R_{n_{\OP{obs}}}=O_{p}(1) \text{ and } g^{\prime }(\tilde{y}_{\OP{mis}})(\hat{y}_{\OP{mis}}-y_{\OP{mis}})=O_{p}(1).
\end{equation*}
\cite{meng2009} claimed that $\hat{v}-v$ is not summarizable because of the \textit{nonnegligiblity} of the remainder term $R_{n_{\OP{obs}}}$, i.e., consistency and asymptotic normality for the MLE $\hat{v}$ from the h-likelihood are not guaranteed. 

Now we investigate the summarizability properties of the MLE $\hat{v}$. In missing data problem, the ML estimation of random parameter can be called the ML imputation. Let $\psi_{0}$ be the true value of $\psi$. As MLE $\hat{\psi}$ is estimating $\psi_{0}$ and similarly the MLE $\hat{y}_{\OP{mis}}$ predicts a realized value of $y_{\OP{mis}}$ by estimating the conditional mode $y_{\OP{mis},0}=\tilde{y}_{\OP{mis}}(\psi _{0}, \mathcal{D}, \boldsymbol{x})$ in (\ref{eqn9}),  which is a function of data and {unknown parameter $\psi_{0}$}. 
This clarifies the summarizability problem raised by \cite{meng2009}; while $\hat{y}_{\OP{mis}} - y_{\OP{mis}}$ is not summarizable, $\hat{y}_{\OP{mis}} - y_{\OP{mis},0}$ is summarizable as in Theorem \ref{thm:consistency_y_mis} below. 
Note that 
\begin{equation*}
    y_{\OP{mis}} - \hat{y}_{\OP{mis}} = y_{\OP{mis},0} - \hat{y}_{\OP{mis}} + \varepsilon,
\end{equation*}
where $\varepsilon =y_{\OP{mis}} - y_{\OP{mis},0}$. In missing data problem, $\varepsilon = O_{p}(1)$. In view of predicting unobservable future (or missing) random variable, we estimate $\varepsilon $ as null. Then, $\hat{y}_{\OP{mis}}$ is estimating $y_{\OP{mis},0}$ to predict $y_{\OP{mis}}$. 
Thus, we obtain 
$$\OP{var}_{\psi} \left(  \hat{y}_{\OP{mis}}-y_{\OP{mis}} \right)= \OP{var}_{\psi}(\hat{y}_{\OP{mis}}-y_{\OP{mis},0})+\OP{var}_{\psi}(\varepsilon|\mathcal{D},\boldsymbol{x}). 
$$
 {The first term is the variance due to estimating $y_{\OP{mis}, 0}$ by $\hat{y}_{\OP{mis}}$ and the second term is the variance due to the unidentifiable error term $\varepsilon$. The second term may decrease with a better imputation model, but  it does not decrease with larger sample size.} 
Moreover, to obtain a standard error for prediction of $y_{\OP{mis}}$, we need to estimate the conditional variance of $\varepsilon$ by using 
\[
\OP{var}_{\psi}(\varepsilon \mid \mathcal{D},\boldsymbol{x})=\OP{var}_{\psi}( y_{\OP{mis}} -  y_{\OP{mis},0} \mid \mathcal{D},\boldsymbol{x})=\OP{var}_{\psi}(y_{\OP{mis}} \mid \mathcal{D},\boldsymbol{x}).
\]
Here, we are interested in estimating $\V(\hat{y}_{\OP{mis}} - y_{\OP{mis}})$. Thus, we write the h-likelihood with respect to $y_{\OP{mis}}$ as $h = h ( \psi ,y_{\OP{mis}} ) =h \{ \psi ,g^{-1}(v) \}$. Note that 
\[
\frac{\partial \tilde{y}_{\OP{mis}}^{\OP{T}}}{\partial \psi }=-I_{\psi
y_{\OP{mis}}}I_{y_{\OP{mis}}y_{\OP{mis}}}^{-1} 
\]
and the variance estimator of $\hat{\psi}$ is $\hat{I}^{\psi \psi }$ by
Theorem \ref{thm:hessian_identity}, where $I_{\psi y_{\OP{mis}}}=-\partial
^{2}h/\partial \psi \partial y_{\OP{mis}}^{\OP{T}}|_{y_{\OP{mis}}=%
\tilde{y}_{\OP{mis}}}$ and $I_{y_{\OP{mis}}y_{\OP{mis}}}=-\partial
^{2}h/\partial y_{\OP{mis}}\partial y_{\OP{mis}}^{\OP{T}}|_{y_{\OP{%
mis}}=\tilde{y}_{\OP{mis}}}$. Then, by using the delta method, we have the
asymptotic normality of $\hat{y}_{\OP{mis}}$ as follows.

\begin{thm}
\label{thm:consistency_y_mis} Under regularity conditions in Appendix, we have 
\[
\sqrt{n_{\OP{obs}}}\left( \hat{y}_{\OP{mis}}-y_{\OP{mis},0}\right) 
\stackrel{\OP{d}}{\rightarrow }\OP{N}\left( 0,V\right) ,
\]
where $V=\lim_{n_{\OP{obs}}\rightarrow \infty }n_{\OP{obs}}\hat{I}_{y_{%
\OP{mis}}y_{\OP{mis}}}^{-1}\hat{I}_{y_{\OP{mis}}\psi }\hat{I}^{\psi
\psi }\hat{I}_{\psi y_{\OP{mis}}}\hat{I}_{y_{\OP{mis}}y_{\OP{mis}%
}}^{-1}$ and $\hat{I}_{\psi y_{\OP{mis}}}$, $\hat{I}_{y_{\OP{mis}}y_{%
\OP{mis}}}$ are evaluated at $\psi =\hat{\psi}$. The variance of $\hat{y}_{%
\OP{mis}}-y_{\OP{mis},0}$ can be estimated as 
\begin{equation}
\widehat{\OP{var}}\left( \hat{y}_{\OP{mis}}-y_{\OP{mis},0}\right) =\OP{var}_{\hat{\psi}}\left( \hat{y}_{\OP{mis}}-y_{\OP{mis},0}\right) =\hat{I}_{y_{\OP{mis}}y_{\OP{mis}}}^{-1}\hat{I}_{y_{\OP{mis}}\psi }%
\hat{I}^{\psi \psi }\hat{I}_{\psi y_{\OP{mis}}}\hat{I}_{y_{\OP{mis}}y_{\OP{mis}}}^{-1}.  \label{eq:var_est_y_mis}
\end{equation}
\end{thm}
{
\noindent If $\E_{\psi}(\varepsilon) = 0$, $\hat{y}_{\OP{mis}}$ is an asymptotically unbiased estimator of $y_{\OP{mis}}$. However, the assumption $\E_{\psi}(\varepsilon) = 0$ is coming from model assumption which may not be identifiable by observed data.} Now, to discuss the estimation of the variance due to the model error $\varepsilon$, suppose that there exists a normalizing transformation $%
z=k(v)=k\{g(y_{\OP{mis}})\}=k\circ g(y_{\OP{mis}})=r(y_{\OP{mis}})$
with $r(\cdot )=k\circ g(\cdot )$ such that $L_{p} (z|  \mathcal{D}; \psi)$ is from the normal density with mean $\tilde{z} = \OP{arg} \max_{z} L_{p}(z|\mathcal{D}; \psi)$ and covariance matrix $I_{zz}^{-1}$, where $I_{zz} = - \partial^{2} h(\psi, z)/ \partial z \partial z^{\T} |_{z = \tilde{z}}$. Then, it gives the h-likelihood 
\[
h\left( \psi ,z\right) =\ell_{m} (\psi )+\frac{1}{2}\log \left| \frac{1}{2\pi }I_{zz}\right| -\frac{1}{2}\left( z-\tilde{z}\right) ^{\OP{T}}I_{zz}\left(
z-\tilde{z}\right).
\]
Here, $\tilde{z}= \E_{\psi} (z |  \mathcal{D}, \B{x}) = r(\tilde{y}_{\OP{mis}})$ provided by the normality of the predictive likelihood $L_{p}(z|\mathcal{D}; \psi)$. This leads to $\E_{\psi}(\varepsilon) = \E_{\psi}(z - z_{0}) = 0$,
\[
\OP{var}_{\psi}\left( \hat{z}-z\right) =\OP{var}_{\psi}\left( \hat{z}-z_{0}\right) +%
\OP{E}_{\psi}\left\{ \OP{var}_{\psi}\left( z_{0}-z\mid \mathcal{D},\boldsymbol{x} \right) \right\} 
\]
and $\widehat{\OP{var}}(z_{0}-z|\mathcal{D},\boldsymbol{x})=\hat{I}_{zz}^{-1}$, where $\hat{z}=r(\hat{y}_{\OP{mis}})$ and $z_{0}=r(y_{\OP{mis},0})=\OP{E}_{\psi_{0}}(z | \mathcal{D},\boldsymbol{x})$. This gives 
\begin{equation*}
\widehat{\OP{var}}\left( \hat{z}-z\right)= \widehat{\OP{var}}\left( 
\hat{z}-z_{0}\right) +\widehat{\OP{var}}\left( z_{0}-z\mid \mathcal{D}, \boldsymbol{x}\right)=\hat{I}_{zz}^{-1}\hat{I}_{z\psi }\hat{I}^{\psi \psi }\hat{I}_{\psi z}%
\hat{I}_{zz}^{-1}+\hat{I}_{zz}^{-1}=\hat{I}^{zz}.
\end{equation*}
Therefore, if a normalizing transformation exists, the h-likelihood gives not only MLEs of both fixed and random parameters, but also their corresponding variance estimators. Moreover, if $y_{\OP{mis}}$ itself satisfies normal approximation well, then, we can have a reasonable variance estimator from the Hessian matrix of h-likelihood 
\begin{eqnarray*}
\widehat{\OP{var}}\left( \hat{y}_{\OP{mis}}-y_{\OP{mis}}\right)  &=&%
\widehat{\OP{var}}\left( \hat{y}_{\OP{mis}}-y_{\OP{mis},0}\right) +%
\widehat{\OP{var}}\left( y_{\OP{mis},0}-y_{\OP{mis}}\mid \mathcal{D},%
\boldsymbol{x}\right)  \\
&=&\hat{I}_{y_{\OP{mis}}y_{\OP{mis}}}^{-1}\hat{I}_{y_{\OP{mis}}\psi }%
\hat{I}^{\psi \psi }\hat{I}_{\psi y_{\OP{mis}}}\hat{I}_{y_{\OP{mis}}y_{%
\OP{mis}}}^{-1}+\hat{I}_{y_{\OP{mis}}y_{\OP{mis}}}^{-1}=\hat{I}^{y_{%
\OP{mis}}y_{\OP{mis}}}.
\end{eqnarray*}
Thus, $\hat{y}_{\OP{mis}, i} \pm 1.96 \sqrt{\hat{I}_{ii}^{y_{\OP{mis}} y_{\OP{mis}} } }$ is 95\% predictive interval of $y_{\OP{mis}, i}$, where $\hat{I}_{ii}^{y_{\OP{mis}} y_{\OP{mis}} }$ is the $i$th diagonal element of $\hat{I}^{y_{\OP{mis}} y_{\OP{mis}} }$. The length of predictive interval is $O_{p}(1)$ and coverage probability becomes exact as $n \to \infty$ \citep{leekim2016}. However, in practice, the normalizing transformation is not known. Thus, in general, for the prediction of $y_{\OP{mis}}$, \cite{leekim2016, leeykimg2020} proposed to use the predictive distribution after eliminating $\psi $ defined as 
\begin{equation}
f(y_{\OP{mis}}\mid \mathcal{D},\boldsymbol{x})=\int f_{\psi }(y_{\OP{mis}%
}\mid \mathcal{D},\boldsymbol{x})c(\psi )d\psi ,
\label{eq:marginal_predictive_lik}
\end{equation}
where $c(\psi )$ is the confidence density \citep{schweder2016}. By using the predictive
likelihood (\ref{eq:marginal_predictive_lik}), we can account for the uncertainty caused by estimating $\psi$. Via simulation studies, \cite{leekim2016} showed that resulting predictive interval maintains the stated coverage probability well as $n$ grows.

\begin{figure}[h!]
    \centering
    \includegraphics[scale=0.4]{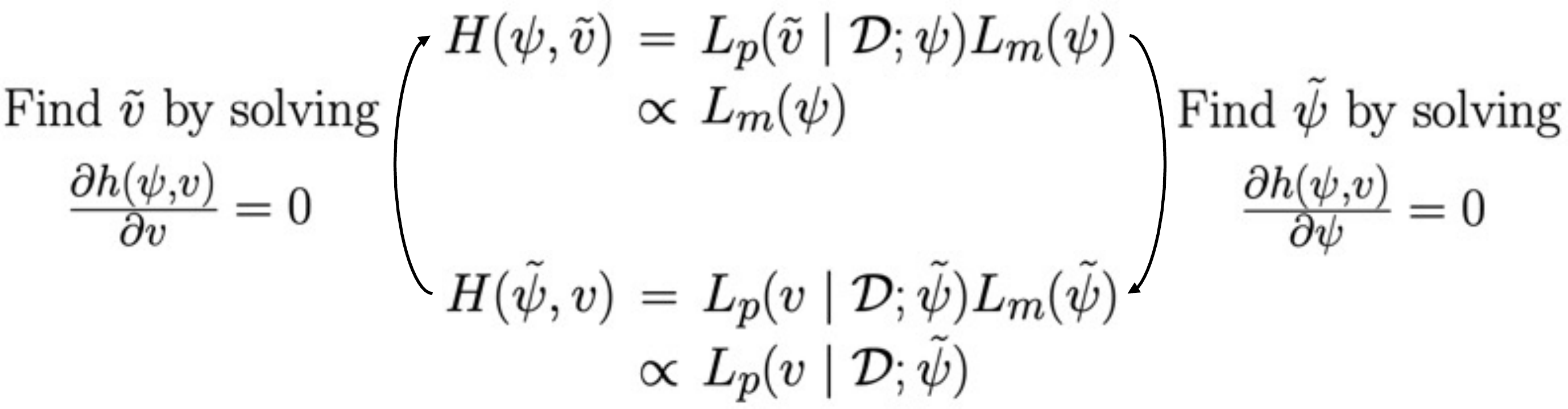}
    \caption{Estimation procedure of the h-likelihood.}
    \label{fig:hlik_procedure}
\end{figure}

From Theorem \ref{thm:hessian_identity}, MLE $\hat{\psi}$ from the marginal likelihood can be obtained by
\begin{equation*}
    \frac{\partial \ell_{m}(\psi)}{\partial \psi} =  \frac{\partial h(\psi, \tilde{v} )}{\partial \psi} = 0
\end{equation*}
and ML imputation $\hat{y}_{\OP{mis}} = g^{-1}(\hat{v})$ of $y_{\OP{mis}} = g^{-1}(v)$ from the predictive likelihood can be obtained by
\begin{equation*}
    \frac{\partial \ell_{p}(v \mid \mathcal{D}; \tilde{\psi})}{\partial v} =  \frac{\partial h(\tilde{\psi}, v)}{\partial v}  = 0,
\end{equation*}
where $\tilde{\psi}$ is solution to $\partial h (\psi, v) / \partial  \psi = 0$. In contrast to the EM algorithm, the h-likelihood provides not only the ML estimation for fixed parameters from $H(\psi, \tilde{v})$, but also ML imputation on random parameters from $H(\tilde{\psi}, v)$ as in Figure \ref{fig:hlik_procedure}. Moreover, the necessary standard error estimates are also given straightforwardly.

\begin{exmp}
\label{exmp:meng} Suppose that $n$ variables are generated from the
exponential distribution with mean $\theta_{0}$ but only the first $n-1$ variables are observed, i.e., $n_{\OP{obs}}=n-1$ and $y_{\OP{mis}} = y_{n}$
is not observed. In this example, the
extended likelihood defined on $y_{\OP{mis}}$-scale is 
\[
\ell _{e}(\theta ,y_{\OP{mis}})=-n\log \theta -\frac{(n-1)\bar{y}_{\OP{%
obs}}+y_{\OP{mis}}}{\theta }.
\]
Note that $y_{\OP{mis}}$-scale is not canonical but $v=\log y_{\OP{mis}}$ is a canonical scale which gives 
\begin{equation*}
h(\theta ,v)=\ell _{e}(\theta ,y_{\OP{mis}})+\log \left| \frac{\partial y_{%
\OP{mis}}}{\partial v}\right| =-n\log \theta -\frac{(n-1)\bar{y}_{\OP{obs%
}}+e^{v}}{\theta }+v  
\end{equation*}
and 
\[
h(\theta ,y_{\OP{mis}})=-n\log \theta -\frac{(n-1)\bar{y}_{\OP{obs}}+y_{%
\OP{mis}}}{\theta }+\log y_{\OP{mis}}.
\]
Here, the canonical function of $y_{\OP{mis}}$ is $\tilde{y}_{\OP{mis}}=\theta $ which gives the MLE $\hat{\theta}=\bar{y}_{\OP{obs}}$ and ML imputation $\hat{y}_{\OP{mis}}=\hat{\theta}=\bar{y}_{\OP{obs}}$. In this example, the MLE of $\theta$, $\hat{\theta}$, satisfies the asymptotic normality 
\[
\sqrt{n_{\OP{obs}}}\left( \hat{\theta}-\theta _{0}\right) \stackrel{\OP{d%
}}{\rightarrow }\OP{N}\left( 0,\theta _{0}^{2}\right) .
\]
By Theorem \ref{thm:consistency_y_mis}, the ML imputation for $y_{\OP{mis}}$, $\hat{y}_{\OP{mis}}$, satisfies the asymptotic normality
\[
\sqrt{n_{\OP{obs}}}\left( \hat{y}_{\OP{mis}}-y_{\OP{mis},0}\right) 
\stackrel{\OP{d}}{\rightarrow }\OP{N}\left( 0,\theta _{0}^{2}\right) ,
\]
where $y_{\OP{mis},0}=\theta _{0}$ and 
$\widehat{\OP{var}}(\hat{y}_{\OP{mis}}-y_{\OP{mis},0})=n_{\OP{obs}%
}^{-1}\hat{\theta}^{2}$, i.e., (\ref{eq:var_est_y_mis}) gives valid variance
estimator of $\hat{y}_{\OP{mis}}-y_{\OP{mis},0}$.
Moreover, 
\[
\hat{I}^{y_{\OP{mis}}y_{\OP{mis}}}=\hat{\theta}^{2}\left( 1+\frac{1}{n_{\OP{obs}}}\right) = \widehat{\OP{var}}(\hat{y}_{\OP{mis}}-y_{\OP{mis}}).
\]
Here $\widehat{\V}(\hat{y}_{\OP{mis}}-y_{\OP{mis}})= \widehat{\V} \left( \hat{y}_{\OP{mis}}-y_{\OP{mis},0}\right)
+ \widehat{\V} (y_{\OP{mis}}|y_{\OP{obs}}) = \hat{\theta}^{2}/n_{\OP{obs}} + \hat{\theta}^{2}$. Thus, the h-likelihood gives a correct ML imputation. In this example, $\tilde{y}_{\OP{mis}, 0} = \theta$ is a function of parameter only so that $\hat{y}_{\OP{mis}} - y_{\OP{mis}, 0}$ is summarizable. But $y_{\OP{mis}}$ is not identifiable since $\varepsilon = O_{p}(1)$ with $\E_{\theta}(\varepsilon) = 0$. Asymptotically correct probability statement on $y_{\OP{mis}}$ can be made based on predictive interval whose length is $O_{p}(1)$.
\end{exmp}

\begin{exmp}
\label{exmp:mixed} Consider a one-way mixed model 
\[
y_{ij}=\mu +u_{i}+\epsilon _{ij},~i=1,\ldots ,q,~j=1,\ldots ,n,
\]
where random effects $u_{i}$ are iid $\OP{N}(0,\lambda ^{2})$, $\epsilon _{ij}$ are iid $\OP{N}(0,\sigma ^{2})$ and $u_{i}$ and $\epsilon _{ij}$ are independent. Henderson's \citet{henderson1959} joint likelihood is the current h-likelihood of 
\cite{lee1996} 
\begin{equation}
\ell _{e}\left( \theta ,u\right) =\sum_{i,j}\left\{ -\frac{1}{2}\log 2\pi
\sigma ^{2}-\frac{1}{2\sigma ^{2}}\left( y_{ij}-\mu -u_{i}\right)
^{2}\right\} +\sum_{i}\left( -\frac{1}{2}\log 2\pi \lambda ^{2}-\frac{1}{%
2\lambda ^{2}}u_{i}^{2}\right) ,  \label{eq:henderson}
\end{equation}
where $\theta =(\mu ,\sigma^{2},\lambda^{2})$. However, joint maximization of (\ref{eq:henderson}) cannot give the MLEs of variance components $\sigma^{2}$ and $\lambda^{2}$. Consider a $v$-scale 
\[
v_{i}=\left\{ -\frac{\partial ^{2}\ell _{e}(\theta ,u)}{\partial u_{i}^{2}}%
\right\} ^{0.5}u_{i}=\left( \frac{\sigma ^{2}+n\lambda ^{2}}{\sigma
^{2}\lambda ^{2}}\right) ^{0.5}u_{i},
\]
which leads to the extended likelihood 
\begin{eqnarray*}
\ell _{e}(\theta ,v) &=&\ell _{e}(\theta ,u)+\log \left| \frac{\partial u}{%
\partial v}\right| =-\frac{N-q}{2}\log 2\pi \sigma ^{2}-\frac{q}{2}\log 2\pi
\left( \sigma ^{2}+n\lambda ^{2}\right)  \\
&&-\frac{1}{2\sigma ^{2}}\sum_{i,j}\left\{ y_{ij}-\mu -\left( \frac{\sigma
^{2}\lambda ^{2}}{\sigma ^{2}+n\lambda ^{2}}\right) ^{0.5}v_{i}\right\} ^{2}-%
\frac{\sigma ^{2}}{2(\sigma ^{2}+n\lambda ^{2})}\sum_{i}v_{i}^{2}-\frac{q}{2}%
\log 2\pi ,
\end{eqnarray*}
where $N=qn$. Since $\ell _{e}( \theta ,\tilde{v}) =\ell_{m} (\theta )$, where 
\begin{equation*}
\tilde{v}_{i} = \tilde{v}_{i} ( \theta ,y_{i} ) = \frac{n\lambda ^{2} ( \bar{y}_{i\cdot}-\mu)}{ \{ \sigma ^{2}\lambda ^{2} ( \sigma ^{2}+n\lambda ^{2} )  \} ^{0.5}},    
\end{equation*}
$y_{i}=(y_{i1},\ldots ,y_{in})$ and $\bar{y}_{i\cdot}=n^{-1}\sum_{j=1}^{n}y_{ij}$, we have h-likelihood $h=\ell _{e}(\theta ,v)$, whose simple maximization gives MLEs of the whole parameters $\theta$.
Also, it gives the best linear unbiased predictors for realized but unobserved random parameters 
\[
\hat{u}_{i}=\tilde{u}_{i}\left( \hat{\theta},y_{i}\right) =\widehat{\OP{E} \left( u_{i}\mid y_{i}\right) },~i=1,\ldots, q,
\]
where 
\[
\tilde{u}_{i}\left( \theta ,y_{i}\right) =\left\{ \frac{\sigma ^{2}\lambda^{2}}{\sigma ^{2}+n\lambda ^{2}}\right\} ^{\frac{1}{2}}\tilde{v}_{i}\left( \theta ,y_{i}\right) =\frac{n\lambda ^{2}}{\sigma ^{2}+n\lambda ^{2}}(\bar{y}%
_{i\cdot }-\mu )=\OP{E}_{\theta} (u_{i}\mid y_{i}).
\]
In this example, the target of $\hat{u}_{i}$ is
\begin{equation*}
  u_{i0} = \tilde{u}_{i}(\theta_{0}, y_{i}) = \OP{E}_{\theta_{0}} (u_{i}\mid y_{i}),  
\end{equation*}
where  $\theta_{0}=(\mu _{0},\sigma _{0}^{2},\lambda _{0}^{2})$ is the true value of $\theta$. If the MLE $\hat{\theta}$ converges to $\theta_{0}$, 
\begin{equation*}
    \widehat{\V} \left ( \hat{u}_{i} - u_{i} \right ) = \widehat{\V} \left ( \hat{u}_{i} - u_{i0} \right ) + \widehat{\V} \left ( u_{i} - u_{i0} \mid y_{i} \right ) \overset{\OP{P}}{\to} 0
\end{equation*}
as $(q, n) \to \infty$. Thus, in this example, we have a consistent estimator of unobserved random parameter $u_{i}$, i.e., $u_{i}$ is identifiable with $\varepsilon = u_{i} - u_{i 0} = o_{p}(1)$. This can also be shown that
\begin{equation*}
\lim_{(q, n) \to \infty} \hat{u}_{i} =\lim_{(q, n) \to \infty }u_{i0} =\lim_{(q, n) \to \infty }\frac{n\lambda _{0}^{2}}{\sigma _{0}^{2}+n\lambda _{0}^{2}}(\bar{y}_{i \cdot }-\mu_{0})=\lim_{(q, n) \to \infty }\frac{n\lambda _{0}^{2}}{\sigma_{0}^{2}+n\lambda _{0}^{2}}\left( u_{i}+\bar{\epsilon}_{i\cdot }\right)
=u_{i},
\end{equation*}
where $\bar{\epsilon}_{i\cdot}=n^{-1}\sum_{j=1}^{n}\epsilon_{ij}$. Model assumptions on $u_{i}$ can also be checkable: for various model checking plots, see \cite{leenelderpawitan2017}. Furthermore, if different model assumptions on $f_{\psi}(u)$ lead to an identical h-likelihood, then it leads to equivalent inferences for identifiable random effects \citep{leenelder2006}. In missing data problem with $\varepsilon = y_{\OP{mis}} - y_{\OP{mis}, 0} = O_{p}(1)$, model assumptions $f_{\psi}(y_{\OP{mis}}  |  \mathcal{D}, \B{x} )$ cannot be checkable from the observed data \citep{molenberghs2008}. 

Since $u_{i}$ itself is the normalizing transformation in this example, variances can be estimated as
\begin{eqnarray*}
\hat{I}_{u_{i} u_{i}}^{-1} &=& \left ( - \frac{
\partial^{2} h}{\partial u_{i}^{2}} \right )^{-1} \Big |_{\theta =  \hat{\theta}} = \frac{\hat{\sigma}^{2} \hat{\lambda}^{2}}{\hat{\sigma}^{2} + n \hat{\lambda}^{2}} = \widehat{\OP{var}} \left ( u_{i} \mid y_{i} \right ) = \widehat{\OP{var}} \left ( u_{i} - u_{i0} \mid y_{i} \right ), \\
\hat{I}^{u_{i} u_{i}} &=& \hat{I}_{u_{i} u_{i}}^{-1} + \frac{\partial \tilde{u}_{i}}{\partial \theta^{\T}} \widehat{\V} \left ( \hat{\theta} \right ) \frac{\partial \tilde{u}_{i}}{\partial \theta} \Big |_{\theta = \hat{\theta}} = \widehat{\V} \left ( \hat{u}_{i} - u_{i} \right ), \\
\hat{I}^{\theta \theta} &=& \widehat{\V} \left ( \hat{\theta} \right ).
\end{eqnarray*}
Thus, proper MLEs of both fixed and random parameters and their variance estimators can be obtained by the maximization of the newly defined h-likelihood, which differs from the joint likelihood of Henderson \citet{henderson1959}. Asymptotically correct probability statement on $u_{i}$ can be made from the predictive interval whose length is $o_{p}(1)$.
For more details about general random effect models, see \cite{paik2015}, \cite{leenelderpawitan2017}, and \cite{leeykimg2020}. 
\end{exmp}

\section{Scale for Joint Maximization}
\label{section:ML_canonicalization}

When the canonical scale is unknown, \cite{leenelderpawitan2017} proposed the use of the
Laplace approximation to give an approximate MLE \citep{tierney1986}, which has been implemented by various packages
\citep{kristensen2016, frailtyHL2019}. In this section, we study how to form an h-likelihood with a weak canonical scale whose joint maximization provides approximate MLEs obtained by the Laplace approximation. Given $y_{\OP{mis}}$-scale, consider a $b$-scale
with $b = g_{1}(y_{\OP{mis}})$. Let $\Omega _{b}$ be the support of $b$
taking a rectangle form $\Omega _{b}=\prod_{i=n_{\OP{obs}%
}+1}^{n}[l_{i},u_{i}]$, where $l_{i}$ and $u_{i}$ are permitted to take the
value of $-\infty $ and $\infty$ with boundary set $\partial \Omega _{b}$, $\xi =(\psi ,b)$ and $f_{\psi }(b)$ be the density function of $b$. \cite{meng2009} studied the regularity conditions for the first and second Bartlett identities of an extended likelihood $\ell _{e}(\psi ,b)$.

\begin{thm}[Meng, 2009]
\label{thm:Bartlett} (i) If $f_{\psi }(b)=0$ for any $b \in \partial \Omega _{b}$, the first Bartlett identity holds.
\begin{equation}
\OP{E}_{\psi}\left[ \frac{\partial }{\partial \xi }\ell _{e}(\psi ,b)\right] =0.  \label{eq:hBartthm1}
\end{equation}
(ii) Furthermore, if $\partial f_{\psi }(b)/\partial b=0$ for any $b\in \partial \Omega _{b}$, the second Bartlett identity holds. 
\begin{equation}
\label{eq:hBartthm2}
\OP{E}_{\psi}\left[ \left( \frac{\partial }{\partial \xi }\ell _{e}(\psi
,b)\right) \left( \frac{\partial }{\partial \xi }\ell _{e}(\psi ,b)\right) ^{%
\OP{T}}\right] +\OP{E}_{\psi}\left[ \frac{\partial ^{2}}{\partial \xi \partial
\xi ^{\OP{T}}}\ell _{e}(\psi ,b)\right] =O.
\end{equation}
Corollary below gives an easy way of having a $b$-scale to satisfy Bartlett identities.
\end{thm}

\begin{cor}
\label{corollary:Bart_cond} Let $\Omega _{b}=\mathbb{R}^{n_{\OP{mis}}}.$ If $\OP{E}_{\psi}\left( b_{i}\right) <\infty $ for all $i$, the $b$-scale satisfies Bartlett identities.
\end{cor}
The second Bartlett identity (\ref{eq:hBartthm2}) guarantees that the predictive likelihood $L_{p}(b | \mathcal{D};\psi )$ is unimodal with respect to $b$ even though $L_{p}(y_{\OP{mis}} | \mathcal{D};\psi )$ may not be unimodal. From Theorem \ref{thm:canonical_existence}, if we have such an extended likelihood $L_{e}(\psi ,b)$ there exists the canonical scale $v=g(b)$ to form the h-likelihood. But, the explicit form of $g(\cdot )$ {for the canonical scale} may not be known. In this case, we {may consider} an approximation of canonical scale based on the Laplace approximation, which is widely used to obtain an approximate MLE of fixed parameter, $\hat{\psi}^{\OP{Lap}}$ \citep{raudenbush2000, leenelderpawitan2017}.
\begin{defn}
\label{defn:weak_canonical}
Suppose that $b$-scale satisfies the Bartlett identities and $\ell_{e}(\psi, b)$ is the corresponding extended log-likelihood. Now, consider a $w$-scale defined as
\begin{equation}
w=g_{2}(b)=\tilde{\Omega}_{bb}^{\frac{1}{2}}b,  \label{eq:canonicalization}
\end{equation}
where $\tilde{b}=\tilde{b}(\psi ,\mathcal{D},\boldsymbol{x})=\OP{arg}%
\max_{b}\ell _{e}(\psi ,b)$ and $\tilde{\Omega}_{bb}=-\partial ^{2}\ell
_{e}(\psi ,b)/\partial b\partial b^{\OP{T}}|_{b=\tilde{b}}$. Here, we call $w$-scale \textit{weak canonical} and
\begin{equation*}
 H = L_{e}(\psi, w) = L_{e}(\psi, b) \left | \frac{\partial b }{ \partial w } \right | 
\end{equation*}
the h-likelihood with weak canonical scale $w$.
\end{defn}
By the {above} definition, weak canonical scale also satisfies Bartlett identities in (\ref{eq:hBartthm1}) and (\ref{eq:hBartthm2}) since the transformation (\ref{eq:canonicalization}) is linear. Furthermore, we have 
$$\tilde{w}=\tilde{w}(\psi ,\mathcal{D},x)=\OP{arg}\max_{w}L_{e}(\psi ,w) = g_{2}\{\tilde{b}(\psi ,%
\mathcal{D},x)\}$$ since $\tilde{b}$ is the
mode of $L_{e}(\psi ,b)$ and the transformation $g_{2}(\cdot )$ is linear. 
Note that the joint maximization of the h-likelihood with weak canonical scale gives the approximate MLE for $\psi$ based on the Laplace approximation as follows.
\begin{equation*}
\hat{L}_{m}(\psi )=L_{e}\left( \psi ,\tilde{b}\right) \left| \frac{1}{2\pi }\tilde{\Omega}_{bb}\right| ^{-\frac{1}{2}}\propto L_{e}\left( \psi ,\tilde{b}%
\right) \left| \frac{\partial b}{\partial w}\right| _{w=\tilde{w}}=L_{e}\left( \psi ,\tilde{w}\right). 
\end{equation*}
This weak canonical scale does not require the existence of linear predictor. In HGLMs, a scale satisfying additivity in
the linear predictor is called a weak canonical scale \citep{leenelderpawitan2017}, which satisfies Corollary \ref{corollary:Bart_cond}. In Appendix, we show how to compute the standard error estimate of the approximate MLE obtained from $\ell _{e}(\psi ,\tilde{w})=\log L_{e}(\psi ,\tilde{w})$. 

\section{ML Imputation}
\label{section:ML_ML}

In this section, we propose the ML imputation via h-likelihood.
\begin{defn}
With the canonical scale $v_{i}=g(y_{\OP{mis},i})$ and the canonical function $\tilde{v}_{i}(\psi ,\mathcal{D},\boldsymbol{x})$, the ML imputation gives imputed values 
\begin{equation}
\hat{y}_{\OP{mis},i}=g^{-1}(\hat{v}_{i}),~\hat{v}_{i}=\tilde{v}_{i}\left( 
\hat{\psi},\mathcal{D},\boldsymbol{x}\right) .
\label{eq:canonical_imputation}
\end{equation}
\end{defn}

\noindent Theorem \ref{thm:consistency_y_mis} implies that the MLE of a random parameter is a consistent estimator of the canonical function. Based on the ML imputation (\ref{eq:canonical_imputation}), we propose to use the estimator
\begin{equation*}
    \bar{y}_{\OP{ML}} = \frac{1}{n} \left (\sum_{i=1}^{n_{\OP{obs}}} y_{i} + \sum_{i=n_{\OP{obs}}+1}^{n} \hat{y}_{\OP{mis}, i} \right )
\end{equation*}
as an estimator of $\eta = \E(Y)$. If the canonical scale is unknown, the ML imputation based on the weak canonical scale can be used. Weak canonical scale always exists and is known. This scale gives the estimator of $\eta$ as 
\begin{equation*}
\bar{y}_{\OP{ML}}^{\OP{Lap}}=\frac{1}{n}\left( \sum_{i=1}^{n_{\OP{obs}%
}}y_{i}+\sum_{i=n_{\OP{obs}}+1}^{n}\hat{y}_{\OP{mis},i}^{\OP{Lap}%
}\right) ,
\end{equation*}
where $\hat{y}_{\OP{mis}}^{\OP{Lap}}=g^{-1}(\hat{w})$, $\hat{w}=\tilde{w}(\hat{\psi}^{\OP{Lap}},\mathcal{D},x)$ and $g=g_{2}\circ g_{1}$. 
From Theorem \ref{thm:canonical_existence} and the definition of the weak canonical scale (\ref{eq:canonicalization}), we see that the canonical scale is a linear transformation of the weak canonical scale $w$. Given $\psi$, MLEs of random parameters are invariant with respect a linear transformation \citep{leenelder2005} and
\begin{eqnarray*}
\hat{\ell}_{m}(\psi )-\ell_{m} (\psi ) &=&\ell_{p}(y_{\OP{mis}}\mid \mathcal{D};\psi )- \hat{\ell}_{p} (y_{\OP{mis}}\mid \mathcal{D};\psi ) \\
&=&\left\{ \ell_{p}(v\mid \mathcal{D};\psi )+\log \left| \partial v/\partial y_{\OP{mis}}\right| \right\} -\left\{ \hat{\ell}_{p}(v \mid \mathcal{D};\psi )+\log \left| \partial v/\partial y_{\OP{mis}}\right| \right\} \\
&=&\ell_{p}(v\mid \mathcal{D};\psi )-\hat{\ell}_{p}(v\mid \mathcal{D};\psi ).
\end{eqnarray*}
Thus, the ML imputation under weak canonical scale is valid in the sense that 
\[
\hat{y}_{\OP{mis}}^{\OP{Lap}}-\hat{y}_{\OP{mis}}=O_{p}\left( \left| \hat{\psi}^{\OP{Lap}}-\hat{\psi}\right| \right), 
\]
where $\hat{\ell}_{p}(y_{\OP{mis}} | \mathcal{D}; \psi) = \log \hat{L}_{p}(y_{\OP{mis}} | \mathcal{D}; \psi)$ and $\hat{L}_{e}(y_{\OP{mis}} | \mathcal{D}; \psi) =
L_{e}(\psi, y_{\OP{mis}}) / \hat{L}_{m}(\psi)$. Recently, \cite{han2022} developed the enhanced Laplace approximation (ELA) to obtain the MLE $\hat{\psi}$ generally. Thus, the ML imputation can be always implemented even when the canonical scale is not known by using a weak canonical scale from the ELA. Given the MLE $\hat{\psi}$, all the results on the ML imputation in Section \ref{section:ML_MLE_random} hold.

Under missing at random (MAR) of \cite{rubin1976}, the h-likelihood becomes 
\[
h=\log f_{\theta }(y_{\OP{obs}}\mid \boldsymbol{x})+\log f_{\theta }(y_{%
\OP{mis}}\mid \boldsymbol{x})+\log f_{\rho }(\delta \mid \boldsymbol{x})+\log \left| \frac{\partial y_{\OP{mis}}}{\partial v}\right| ,
\]
where $\theta$ is the parameter for the response model and $\rho$ is the parameter associate with the missing mechanism. Under MAR assumption,
the canonical function of $v$ depends only on $\theta $ and $\boldsymbol{x}$
to give ML imputed values $\hat{y}_{\OP{mis},i}=\tilde{y}_{\OP{mis}, i} (\hat{\theta}, \B{x}_{i})$, $\tilde{y}_{\OP{mis}, i}(\theta, \B{x}_{i}) = g^{-1}\{\tilde{v}_{i}(\theta,\boldsymbol{x}_{i})\}$. 

\begin{exmp}
\label{exmp:cem} 
\cite{little2019} considered censored exponential model, where $y_{\OP{com}}=(y_{\OP{obs}},y_{\OP{mis}}) $ are independent exponential random variables with mean $\theta $ and the missing mechanism is set to $\delta =I(Y\leq c)$ with known $c$. Here the missing mechanism is not ignorable and the complete-data likelihood is 
\[
\ell _{e}\left( \theta ,y_{\OP{mis}}\right) =-n\OP{log}\theta -\frac{1}{%
\theta }\sum_{i=1}^{n_{\OP{obs}}}y_{i}-\frac{1}{\theta }\sum_{i=n_{\OP{%
obs}}+1}^{n}y_{\OP{mis},i}.
\]
They noted that joint maximization of the complete-data likelihood provides nonsensical modes $(n_{\OP{obs}} \bar{y}_{\OP{obs}} + n_{\OP{mis}}c)/n$ for $\theta$ and $c$ for $y_{\OP{mis}, i}$, where $\bar{y}_{\OP{obs}} = \sum_{i=1}^{n_{\OP{obs}}} y_{i} / n_{\OP{obs}}$ is the sample mean based on the observed responses. Now we know that MLEs (modes) should be obtained from the h-likelihood. \cite{yun2007} found the canonical scale $v_{i}=\OP{log}\left( y_{\OP{mis},i}-c\right) $ to form the h-likelihood 
\[
h=\ell _{e}\left( \theta ,y_{\OP{mis}}\right) +\OP{log}\left| \frac{\partial y_{\OP{mis}}}{\partial v}\right| =-n\OP{log}\theta -\frac{1}{\theta }\sum_{i=1}^{n_{\OP{obs}}}y_{i}+\sum_{i=n_{\OP{obs}}+1}^{n}\left\{ -\frac{1}{\theta }\left( c+e^{v_{i}}\right) +v_{i}\right\} .
\]
The canonical function of $v$ is $\tilde{v}_{i}(\theta )=\log \theta $ which gives 
\[
h\left\{ \theta ,\tilde{v}(\theta )\right\} =-n_{\OP{obs}}\OP{log}\theta
-\frac{1}{\theta }\sum_{i=1}^{n_{\OP{obs}}}y_{i}-\frac{n_{\OP{mis}}c}{%
\theta }-n_{\OP{mis}}=\ell_{m} (\theta )-n_{\OP{mis}} \propto \ell_{m} (\theta ).
\]
This gives the true MLE $\hat{\theta}=\bar{y}_{\OP{obs}}+n_{\OP{mis%
}}c/n_{\OP{obs}}$ and the ML imputed values $\hat{y}_{\OP{mis},i}=\hat{\theta}+c>c$ to lead that
\begin{equation*}
    \bar{y}_{\OP{ML}} = \frac{1}{n} \left (  \sum_{i=1}^{n_{\OP{obs}}} y_{i} + \sum_{i=n_{\OP{obs}}+1}^{n} \hat{y}_{\OP{mis}, i} \right ) = \hat{\theta}
\end{equation*}
and $\widehat{\OP{var}}(\bar{y}_{\OP{ML}})=\widehat{\OP{var}}(\hat{\theta})=\hat{I}^{\theta \theta }=\hat{\theta}^{2}/n_{\OP{obs}}$. \cite{little2019} used the EM algorithm. With the E-step
\begin{equation*}
\OP{E}_{\theta}(y_{\OP{mis},i}|y_{\OP{mis},i}>c)=\theta +c,    
\end{equation*}
the M-step gives
\begin{equation*}
    \theta^{(t+1)} = \frac{1}{n} \left [ \sum_{i=1}^{n_{\OP{obs}}}y_{i} + n_{\OP{mis}} \left \{ \theta^{(t)} + c \right \} \right ].
\end{equation*}
Thus, the EM algorithm gives the identical MLE $\hat{\theta}$. But, the EM algorithm does not provide the variance estimator directly.

To examine the performance of the ML imputation, we set about 22\% of responses as unobserved and compare three estimators $\bar{y}_{\OP{com}} = \sum_{i=1}^{n}y_{i}/n$, $\bar{y}_{\OP{obs}} = \sum_{i=1}^{n}\delta_{i}y_{i}/n_{\OP{obs}}$, and $\bar{y}_{\OP{ML}}$ using random samples
from $\exp (2)$ distribution. The estimator $\bar{y}_{\OP{com}}$ is considered as a benchmark since it cannot be used in practice. In Figure \ref{fig:emm}, it is shown that the proposed method works well. Moreover, $\bar{y}_{\OP{obs}}$ shows a
non-negligible bias in amount $n_{\OP{mis}}c/n_{\OP{obs}}\approx 0.86$
since the missing mechanism is not ignorable.
\begin{figure}[h!]
\centering
\begin{subfigure}{.4\textwidth}
  \centering
  \includegraphics[scale=0.5]{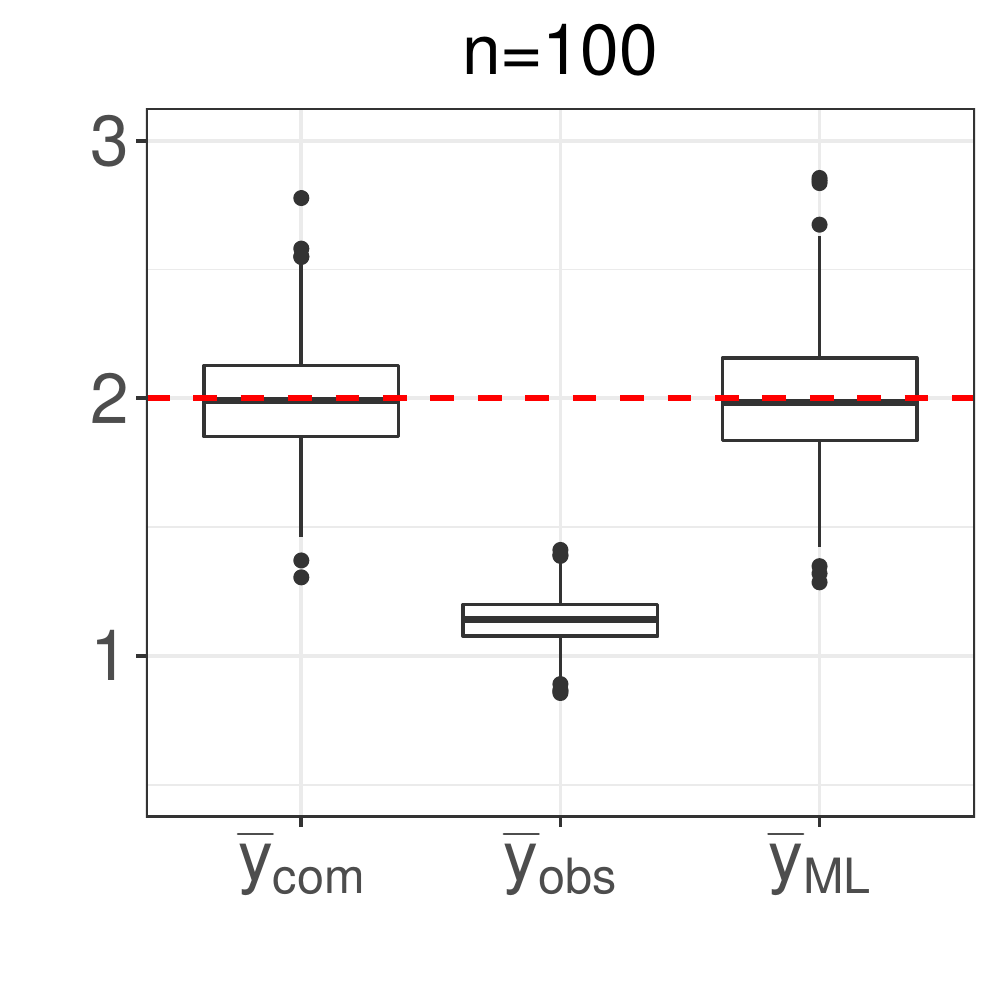}
\end{subfigure}%
\begin{subfigure}{.4\textwidth}
  \centering
  \includegraphics[scale=0.5]{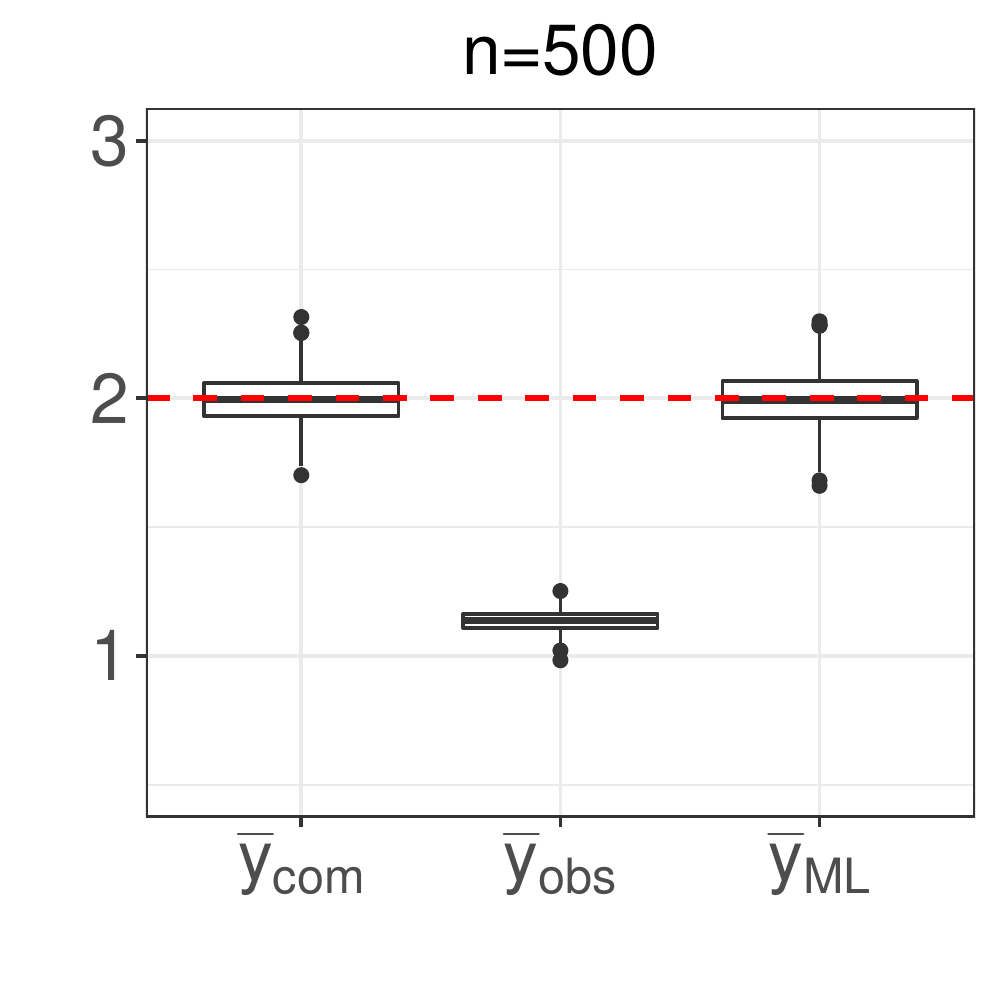}
\end{subfigure}
\caption{Boxplots of estimators in exponential mean model with $c=3$. Dotted line indicates the true value of $\eta$.}
\label{fig:emm}
\end{figure}
\end{exmp}

\section{Illustrative Examples}
\label{section:ML_examples}

\subsection{Normal Regression Model}

\label{subsection:nrm} Consider a normal regression model $Y|x\sim \OP{N}(\beta _{0}+\beta _{1}x,\sigma ^{2})$ with response probability model  $\OP{logit}\{\OP{P}_{\rho}(\delta =1|x)\}=\rho _{0}+\rho _{1}x+\rho _{2}x^{2}$ under a MAR assumption. Here, $y_{\OP{mis}}$-scale itself satisfies the Bartlett identities but it is
canonical scale only for $(\beta _{0},\beta _{1})$. Thus, the joint
maximization of $\ell_{e}(\theta, y_{\OP{mis}})$ cannot give the MLE of $%
\sigma ^{2}$, where $\theta =(\beta _{0},\beta _{1},\sigma ^{2})$. However, $v$-scale defined by $v_{i} = y_{\OP{mis}, i} / \sigma$ is the canonical scale with canonical function $\tilde{v}_{i}(\theta ,x_{i})=(\beta _{0}+\beta _{1}x_{i}) / \sigma$ for $i=n_{\OP{obs}}+1, \ldots ,n$. Then,
the canonical function of $y_{\OP{mis}}$ is $\tilde{y}_{\OP{mis}, i}
(\theta, x_{i}) = \beta_{0} + \beta_{1} x_{i} = \OP{E}_{\theta}(y_{\OP{mis}, i} |
x_{i})$ and the ML imputed values are $\hat{y}_{\OP{mis},i}=\hat{\beta _{0}%
}+\hat{\beta _{1}}x_{i}$. Moreover, 
\[
\hat{I}_{y_{\OP{mis}, i} y_{\OP{mis}, i}}^{-1} = \left ( - \frac{%
\partial^{2} h}{\partial y_{\OP{mis}, i}^{2}} \right )^{-1} \Big |_{\theta
= \hat{\theta}} = \hat{\sigma}^{2} = \widehat{\OP{var}} \left ( y_{\OP{mis}, i} \mid \mathcal{D}, \boldsymbol{x} \right ).
\]
Since
\begin{equation*}
    \tilde{y}_{\OP{mis}, i}
(\theta, x_{i}) = \OP{E}_{\theta}(y_{\OP{mis}, i} |
x_{i}),
\end{equation*}
the MLEs can also be obtained by the EM algorithm.

\begin{figure}[h!]
\centering
\begin{subfigure}{.4\textwidth}
  \centering
  \includegraphics[scale=0.5]{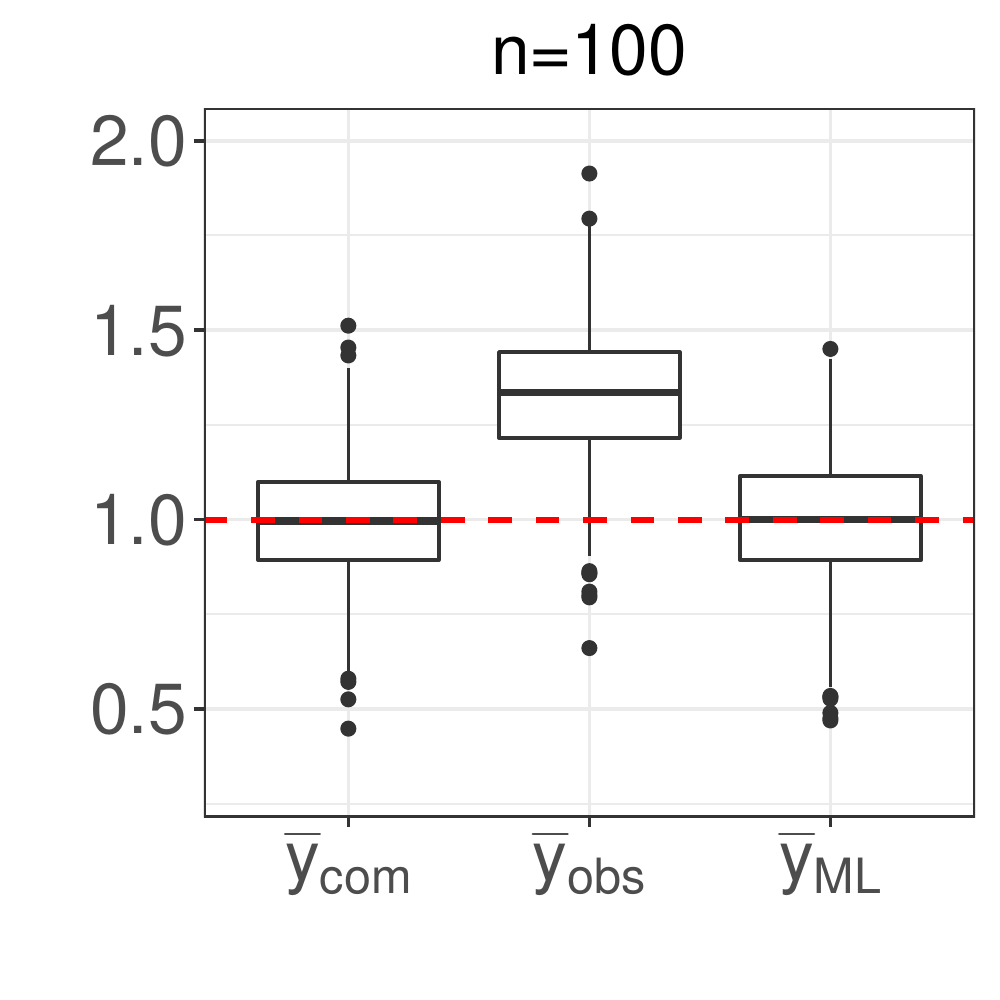}
\end{subfigure}
\begin{subfigure}{.4\textwidth}
  \centering
  \includegraphics[scale=0.5]{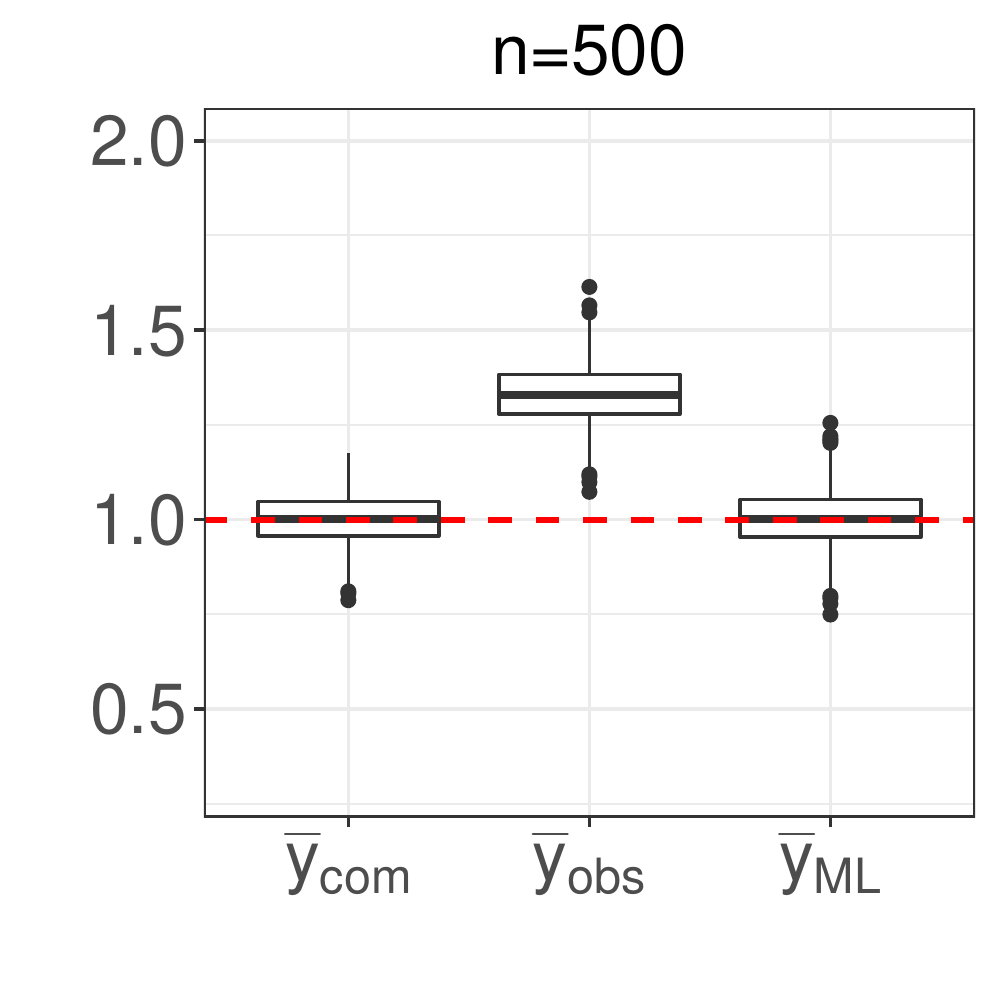}
\end{subfigure}
\caption{Boxplots of estimators in normal regression model. Dotted line indicates the true value of $\eta$.}
\label{fig:nrm}
\end{figure}

For a simulation study, we generate $n=100$ and $n=500$ samples with $\theta =(1,2,1)$, $\rho =(1,2,0.3)$ and $x\sim \OP{U}(-1,1)$%
. From Figure \ref{fig:nrm}, we can see that $\bar{y}_{\OP{obs}}$ is
positively biased because the covariate $x $ increases both $\OP{E}_{\theta}(Y|x)$
and $\OP{P}_{\rho}(\delta =1|x)$. Also, the performance of $\bar{y}_{\OP{ML}}$
is almost same as $\bar{y}_{\OP{com}}$.

\subsection{Exponential Regression Model}

\begin{figure}[h!]
\centering
\begin{subfigure}{.4\textwidth}
  \centering
  \includegraphics[scale=0.5]{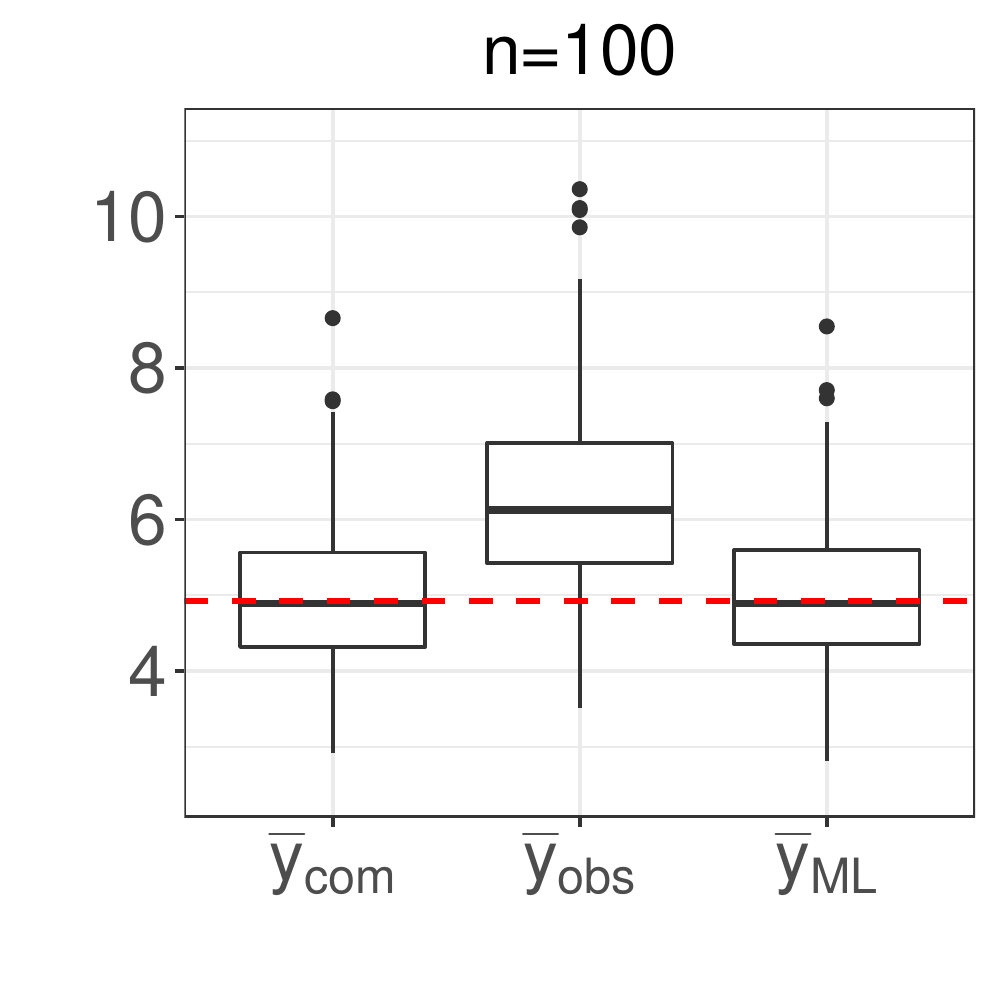}
\end{subfigure}
\begin{subfigure}{.4\textwidth}
  \centering
  \includegraphics[scale=0.5]{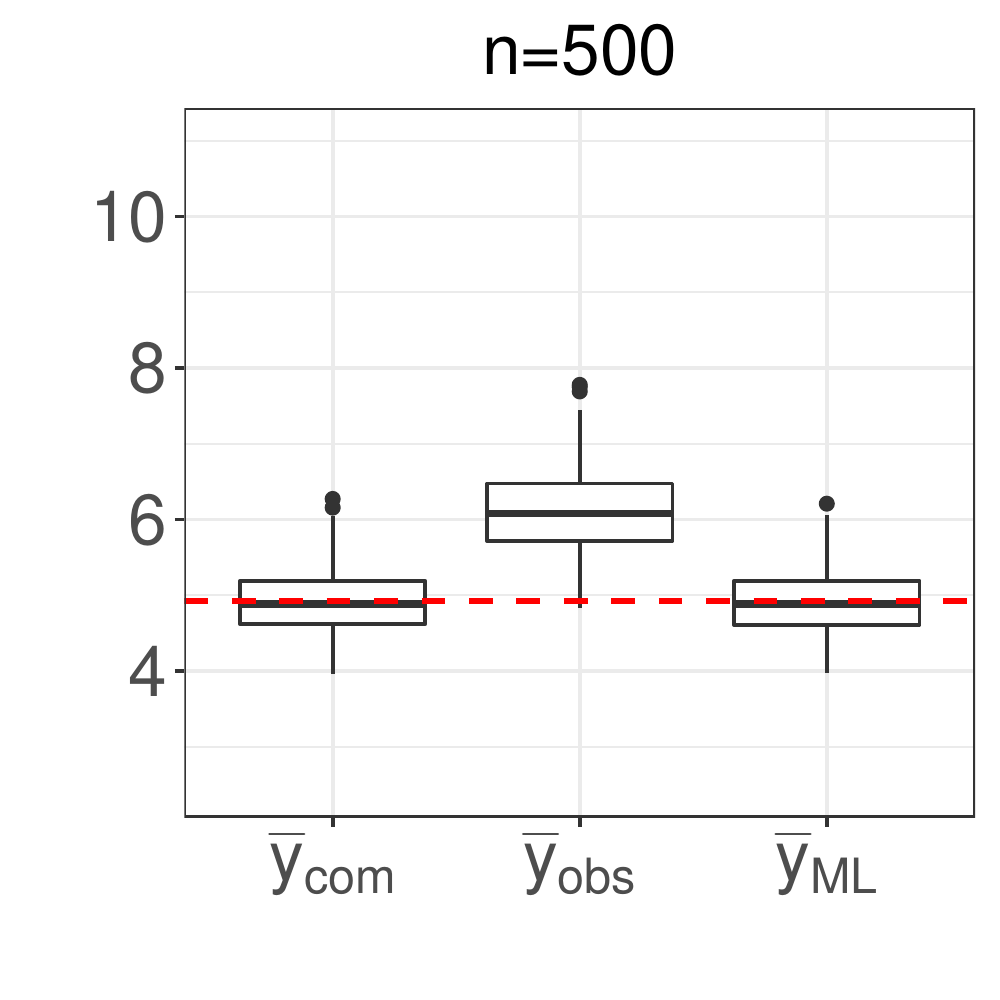}
\end{subfigure}
\caption{Boxplots of estimators in exponential regression model. Dotted line indicates the true value of $\eta$.}
\label{fig:erm}
\end{figure}

Consider an exponential regression model with mean $\OP{E}_{\beta}\left(Y|x\right) =\exp (\beta _{0}+\beta _{1}x)$, $\beta = (\beta_{0}, \beta_{1})$ and the MAR mechanism as the
Example \ref{subsection:nrm}. In this example, $v = \log y_{\OP{mis}}$
scale is the canonical scale which also satisfies Bartlett identities by
Corollary \ref{corollary:Bart_cond}. Here the canonical function of $y_{\OP{mis}, i}$ is $\tilde{y}_{\OP{mis}, i} = \exp (\beta_{0} + \beta_{1}
x_{i}) = \OP{E}_{\beta}(y_{\OP{mis}, i} | x_{i})$ and the ML imputed values are $\hat{y}_{\OP{mis},i}=\exp (\hat{\beta}_{0}+\hat{\beta}_{1}x_{i})$.  Moreover, 
\[
\hat{I}_{y_{\OP{mis}, i} y_{\OP{mis}, i}}^{-1} = \left ( - \frac{%
\partial^{2} h}{\partial y_{\OP{mis}, i}^{2}} \right )^{-1} \Big |_{\theta = \hat{\theta}} = \hat{y}_{\OP{mis}, i}^{2} = \widehat{\OP{var}} \left (
y_{\OP{mis}, i} \mid \mathcal{D}, \boldsymbol{x} \right ).
\]
Figure \ref{fig:erm} shows simulation results with $\beta $ and $\rho $
being the same as in Example \ref{subsection:nrm}. Compared to $\bar{y}_{\OP{com}}$, $\bar{y}_{\OP{ML}}$ gives almost the same performances, whereas $\bar{y}_{\OP{obs}}$ is biased.

\subsection{Tobit Regression Model}

Suppose that responses are generated from the normal regression model in Example \ref{subsection:nrm}. In addition, missing data are created by $y_{\OP{mis}}>c$ at a known censoring point $c$. The extended likelihood 
\[
\ell _{e}\left( \theta ,y_{\OP{mis}}\right) =-\frac{n}{2}\OP{log}2\pi
\sigma ^{2}-\frac{1}{2\sigma ^{2}}\sum_{i=1}^{n_{\OP{obs}}}\left( y_{i}-%
\tilde{x}_{i}^{\OP{T}}\beta \right) ^{2}-\frac{1}{2\sigma ^{2}}\sum_{i=n_{%
\OP{obs}}+1}^{n}\left( y_{\OP{mis},i}-\tilde{x}_{i}^{\OP{T}}\beta
\right) ^{2}, 
\]
where $\theta =(\beta ,\sigma ^{2})$, $\beta =(\beta _{0},\beta _{1})$ and $%
\tilde{x}=(1,x)$. Here a $b$-scale 
\[
b_{i}=g_{1}\left( y_{\OP{mis},i}\right) =\OP{log}\left( y_{\OP{mis},
i} - c \right) , 
\]
satisfies Bartlett identities by Corollary \ref{corollary:Bart_cond} but it
is not canonical. Now, consider a $w$-scale with $w_{i}=g_{2}(b_{i})=\tilde{%
\Omega}_{b_{i} b_{i}}^{0.5}b_{i}$ by (\ref{eq:canonicalization}). Then, we
have the approximate MLE $\hat{\theta}^{\OP{Lap}}$ and approximate ML
imputed values $\hat{y}_{\OP{mis}}^{\OP{Lap}}$ by jointly maximizing $%
\ell _{e}(\theta ,w)$. However, the exact marginal log-likelihood is
available in Tobit regression model. 
\[
\ell_{m} (\theta )=-\frac{n_{\OP{obs}}}{2}\log \sigma ^{2}-\frac{1}{2\sigma
^{2}}\sum_{i=1}^{n_{\OP{obs}}}\left( y_{i}-\tilde{x}_{i}^{\OP{T}}\beta
\right) ^{2}+\sum_{i=n_{\OP{obs}}+1}^{n}\log \left\{ \Phi \left( \frac{%
\tilde{x}_{i}^{\OP{T}}\beta -c}{\sigma }\right) \right\} . 
\]
This means that explicit form of the predictive likelihood $L_{e}(b_{i}|y_{%
\OP{obs}};\theta )$ is available to give the canonical scale 
\begin{equation}
v_{i}=L_{e}\left( \tilde{b}_{i}\mid y_{\OP{obs}};\theta \right) b_{i},
\label{eq:wav}
\end{equation}
where 
\begin{equation*}
\tilde{b}_{i}=\log \left\{ \tilde{x}_{i}^{\OP{T}}\beta -c+\sqrt{\left( 
\tilde{x}_{i}^{\OP{T}}\beta -c\right) ^{2}+4\sigma ^{2}}\right\} -\log 2.
\end{equation*}
Thus, all MLEs are computed directly by simple maximization of the h-likelihood. 

In the simulation study, we examine the performance of ML imputations by using two estimators $\bar{y}_{\OP{ML}}$ using the MLE and $\bar{y}_{\OP{%
MI}}^{\OP{Lap}}$ using the approximate MLE. From (\ref{eq:wav}), we see that both $b$ and $w$ are linear transformations of $v$. Thus, approximate
ML imputation works well as approximate MLE does. Given MLE for fixed parameters, weak canonical scale gives an exact ML imputation.

For simulation, we set $\theta =(1,3,1)$, $c=3$ and $x_{i}=-1+2i/n$ for $%
i=1,\ldots ,n$. 
\begin{figure}[h!]
\centering
\begin{subfigure}{.45\textwidth}
  \centering
  \includegraphics[scale=0.45]{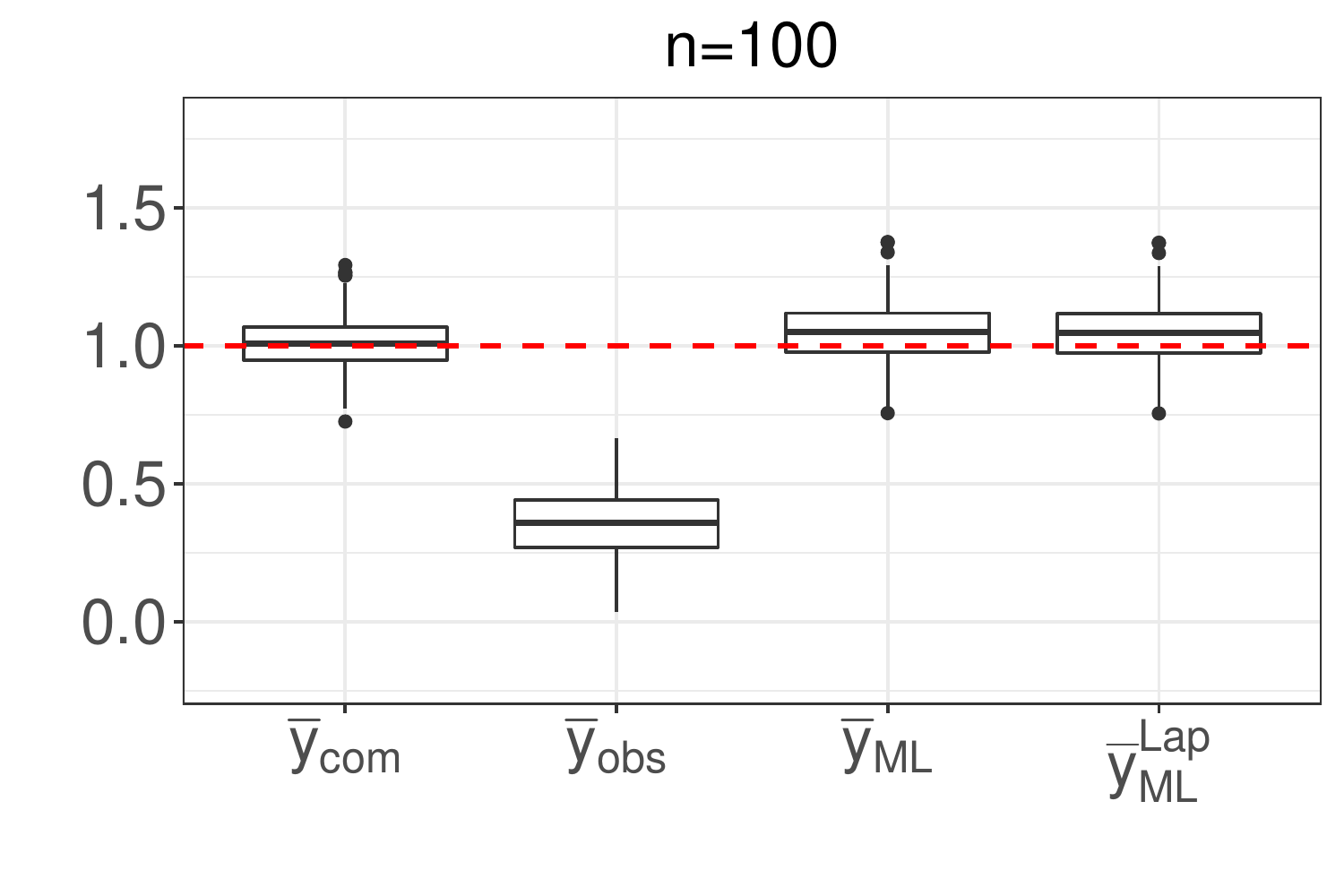}
\end{subfigure}%
\begin{subfigure}{.45\textwidth}
  \centering
  \includegraphics[scale=0.45]{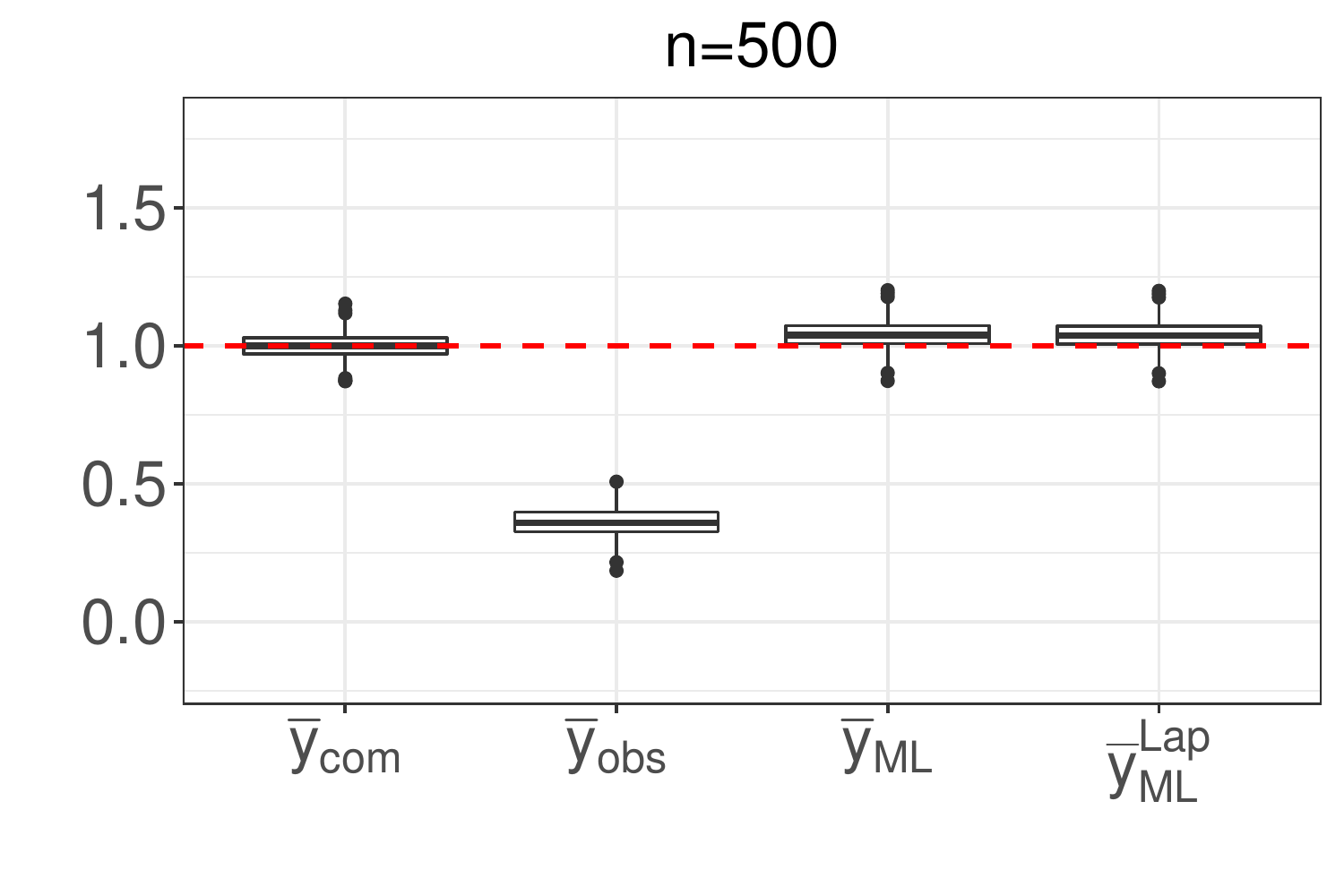}
\end{subfigure}
\caption{Boxplots of estimators in Tobit regression model. Dotted line indicates the true value of $\eta$.}
\label{fig:Tobit}
\end{figure}
In Figure \ref{fig:Tobit}, we see that the difference
between $\bar{y}_{\OP{ML}}$ and $\bar{y}_{\OP{ML}}^{\OP{Lap}}$ is
negligible because $\hat{\theta}$ and $\hat{\theta}^{\OP{Lap}}$ are very
close. Therefore, we can use the weak canonical scale and approximate MLE
when canonical scale is unknown. 

\section{Conclusion}
\label{section:ML_conclusion}

\cite{firth2006} and \cite{meng2009} raised two important reservations about the use of the h-likelihood. \cite{firth2006} noted that the linear predictor in HGLM may not be well-defined to form the h-likelihood. \cite{leenelderpawitan2006} resolved his question by defining the canonical scale. \cite{meng2009} claimed  the asymptotic theory for the  prediction of the future data would be impossible because the consistency cannot be achieved for the predicted values from the h-likelihood. In this paper, we have answered their queries on the h-likelihood in the context of imputation for missing data. Specifically, we have shown  that prediction becomes an estimation of canonical function of the h-likelihood whose consistent estimation and asymptotic normality can be justifiable. We further showed that standard errors of prediction can be directly obtained from the h-likelihood.

\cite{little2019} pointed out that the current h-likelihood procedure achieves the correct ML estimation by modifying h-likelihood. In this paper, we achieve the true ML approach via h-likelihood without any modification by reformulating the h-likelihood. We present the meaning of the canonical scale and canonical function in detail, which allow ML estimation of fixed parameters and ML imputation of random parameters, namely missing data. The Jacobian term is a key to finding the canonical scale.

The ML imputation using the h-likelihood estimates the conditional mode, rather than the conditional mean of the missing value. We call this conditional mode imputation the ML imputation for the random parameters.
The h-likelihood used for ML imputation provides an efficient algorithm because resampling procedure for multiple imputations or expectation steps in EM algorithm is not compulsory.



\bibliographystyle{apalike}
\bibliography{converted_to_latex.bib}


\section*{Appendix: Supplementary Materials for ``Maximum Likelihood Imputation''}

\subsection*{A1 Regularity Conditions}

In this paper, we assume the following regularity conditions in developing
the proposed method.
\begin{itemize}
\item[(R1)]  Let $\psi _{0}=\OP{arg}\OP{max}_{\psi }\OP{E}_{\psi}\{\ell_{m} (\psi
)\}$ be the true value of $\psi$. Here, the number of fixed parameters does not depend on $n_{\OP{obs}}$. Then, the MLE $\hat{\psi} = \OP{arg} \max_{\psi} \ell_{m}(\psi)$ satisfies the asymptotic
normality with mean $\psi_{0}$ and variance $\mathcal{I}_{0}^{-1}=\mathcal{I}%
^{-1}\left( \psi_{0}\right) $, where 
\[
\mathcal{I}\left( \psi \right) =\lim_{n_{\OP{obs}} \to \infty }\frac{1}{n_{%
\OP{obs}}}\left( -\frac{\partial ^{2}\ell_{m} (\psi )}{\partial \psi \partial
\psi ^{\OP{T}}}\right) \Big |_{\psi =\psi_{0}} 
\]
is the expected Fisher information.

\item[(R2)]  The support of missing values 
\[
\Omega _{y_{\OP{mis}}}=\left\{ y_{\OP{mis}}\in \mathbb{R}^{n_{\OP{mis}%
}}:\prod_{i=n_{\OP{obs}}+1}^{n}f_{\psi }\left( y_{\OP{mis},i},\delta
_{i}=0\mid \boldsymbol{x}_{i}\right) >0\right\} \subset \mathbb{R}^{n_{\OP{mis%
}}} 
\]
does not depend on fixed parameter $\psi $.
\end{itemize}

\subsection*{A2 Proofs}
\subsubsection*{A2.1 Proof of Theorem 3.1}

\begin{proof}
By assumption, there exists $\tilde{y}_{\OP{mis}} = \OP{arg} \max_{y_{\OP{mis}}} \ell_{e}(\psi, y_{\OP{mis}})$. Now, consider a $v$-scale defined by
\begin{equation*}
    v_{i} = g(y_{\OP{mis}, i}) = \left \{ L_{p} \left ( \tilde{y}_{\OP{mis}} \mid \mathcal{D}; \psi \right ) \right \}^{1/n_{\OP{mis}}} y_{\OP{mis}, i}, ~ i=n_{\OP{obs}}+1, \ldots, n,
\end{equation*}
with the predictive likelihood $L_{p}(y_{\OP{mis}} \mid \mathcal{D}; \psi) = f_{\psi}(y_{\OP{mis}} \mid \mathcal{D}, \B{x})$. Here, the transformation $g(\cdot)$ is bijective and differentiable since it is linear. The predictive likelihood on $v$-scale is also well-defined with the Jacobian term
\begin{equation*}
    L_{p}(v \mid \mathcal{D}; \psi) = L_{p}(y_{\OP{mis}} \mid \mathcal{D}; \psi)\left| \frac{\partial y_{\OP{mis}}}{\partial v}\right|, ~ \left| \frac{\partial v}{\partial y_{\OP{mis}}}\right| _{v=\tilde{v}%
} = L_{p}(\tilde{y}_{\OP{mis}}\mid \mathcal{D};\psi ),
\end{equation*}
where $\tilde{v}_{i} = g(\tilde{y}_{\OP{mis}, i})$. Note that $\tilde{v}$ is also the mode of $L_{p}(v | \mathcal{D}; \psi)$ since $\tilde{y}_{\OP{mis}}$ is the mode of $L_{p}(y_{\OP{mis}} | \mathcal{D}; \psi)$ and the transformation $g(\cdot)$ is linear. Therefore, there exists a canonical scale which satisfies (12).
\end{proof}

\subsubsection*{A2.2 Proof of Theorem 3.2}

\begin{proof}
Let $v$-scale be the canonical scale and $\tilde{v} = \tilde{v}(\psi, \mathcal{D}, \B{x})$. Then, the h log-likelihood can be written as
\begin{equation*}
h \left ( \psi, \tilde{v} \right ) = \ell_{m} \left ( \psi \right ) + c,
\end{equation*}
where $c$ is a constant which is free of $\psi$. Then, we can prove the first equality
\begin{equation*}
\frac{\partial}{\partial \psi} h \left ( \psi, \tilde{v} \right ) = \frac{\partial h}{\partial \psi} \Big |_{v = \tilde{v}}  + \frac{\partial \tilde{v}^{\T} }{\partial \psi} \frac{\partial h}{\partial v} \Big |_{v = \tilde{v}} = \frac{\partial h}{\partial \psi} \Big |_{v = \tilde{v}} = \frac{\partial \ell_{m}}{\partial \psi},
\end{equation*}
where $h = h(\psi, v)$ and $\ell_{m} = \ell_{m}(\psi)$. To show the second equality, recall that
\begin{equation}
\label{eq:Jacobian}
\frac{\partial h}{\partial v} \Big |_{v = \tilde{v}} = 0.
\end{equation}
By differenciating (\ref{eq:Jacobian}) with respect to $\psi$,
\begin{equation}
\frac{\partial^{2} h}{\partial \psi \partial v^{\T}}  \Big |_{v = \tilde{v}} + \frac{\partial \tilde{v}^{\T} }{\partial \psi} \left \{ \frac{\partial^{2} h}{\partial v \partial v^{\T}} \right \}_{v = \tilde{v}} = O. ~ \Rightarrow ~ \frac{\partial \tilde{v}^{\T} }{\partial \psi} = - I_{\psi v} I_{v v}^{-1} |_{v = \tilde{v}}.
\label{eq:Jacobian_tilde}
\end{equation}
Therefore, from (\ref{eq:Jacobian_tilde}), we can prove the required result.
\begin{eqnarray*}
\frac{\partial \ell_{m}}{\partial \psi} &=& \frac{\partial}{\partial \psi} h \left ( \psi, \tilde{v} \right ). \\
\Rightarrow - \frac{\partial^{2} \ell_{m}}{\partial \psi \partial \psi^{\T}} &=& - \frac{\partial^{2} h}{\partial \psi \partial \psi^{\T}} \Big |_{v = \tilde{v}} - \frac{\partial \tilde{v}^{\T} }{\partial \psi} \frac{\partial^{2} h}{\partial v \partial \psi^{\T}} \Big |_{v = \tilde{v}} \\
&=&  I_{\psi \psi} -  I_{\psi v} I_{v v}^{-1} I_{v \psi} \\
&=&  \left ( I^{\psi \psi} \right )^{-1}.
\end{eqnarray*}
Here,
\begin{eqnarray*}
\begin{pmatrix}
I^{\psi \psi } & I^{\psi v} \\
I^{v\psi } & I^{vv}
\end{pmatrix}
&=&
\begin{pmatrix}
I_{\psi \psi } & I_{\psi v} \\
I_{v\psi } & I_{vv}
\end{pmatrix}
^{-1}, \\
\begin{pmatrix}
I_{\psi \psi } & I_{\psi v} \\
I_{v\psi } & I_{vv}
\end{pmatrix}
&=&
\begin{pmatrix}
-\partial ^{2}h / \partial \psi \partial \psi ^{\OP{T}} & -\partial ^{2}h
/ \partial \psi \partial v^{\OP{T}} \\
-\partial ^{2}h / \partial v\partial \psi ^{\OP{T}} & -\partial ^{2}h /
\partial v\partial v^{\OP{T}}
\end{pmatrix}
_{v=\tilde{v}}.
\end{eqnarray*}
\end{proof}

\subsubsection*{A2.3 Proof of Corollary 4.1}

\begin{proof}
It suffices to show that the case $n_{\OP{mis}}=1$. If $\E_{\psi}(b) < \infty$, then
\begin{equation*}
\lim_{|b| \to \infty} b f_{\psi} (b, \delta=0 \mid \B{x}) = 0 ~ \Rightarrow ~ \lim_{|b| \to \infty} f_{\psi} (b, \delta=0 \mid \B{x}) = 0.
\end{equation*}
Since $f_{\psi}$ is continuous, $f_{\psi} (b, \delta=0 \mid \B{x}) = 0$ for $b \in \partial \Omega_{b} = \left \{ -\infty, \infty \right \}$. Moreover, $f_{\psi}$ is bounded since $f_{\psi}$ is a density function of a continuous random variable whose support is the whole real line with finite mean. This guarantees that $f_{\psi}$ is uniformly continuous which implies
\begin{eqnarray*}
\lim_{|b| \to \infty} f_{\psi}^{\prime} (b, \delta=0 \mid \B{x}) &=& \lim_{|b| \to \infty} \lim_{t \to 0} \frac{f_{\psi} (b + t, \delta=0 \mid \B{x}) - f_{\psi}(b, \delta=0 \mid \B{x})}{t} \\
&=& \lim_{t \to 0} \lim_{|b| \to \infty} \frac{f_{\psi} (b + t, \delta=0 \mid \B{x}) - f_{\psi}(b, \delta=0 \mid \B{x})}{t} \\
&=& 0,
\end{eqnarray*}
i.e., $f_{\psi}^{\prime} (b, \delta=0 \mid \B{x}) = 0$ for $b \in \partial \Omega_{b} = \left \{ -\infty, \infty\right \}$. Then, the first and second Bartlett identities hold by the result of Theorem 4.1.
\end{proof}

\subsubsection*{A2.4 Score and Hessian of $\hat{\ell}_{m} (\protect\psi)$ and $\ell_{e} ( 
\protect\psi, \tilde{w} )$}

By the definition of $\hat{\ell}_{m}(\psi)$, the score and Hessian can be expressed as 
\begin{eqnarray*}
\frac{\partial}{\partial \psi_{j}} \hat{\ell}_{m} (\psi) &=& \frac{\partial}{%
\partial \psi_{j}} \ell_{e} (\psi, b) \Big |_{b = \tilde{b} } - \frac{1}{2} 
\OP{tr} \left \{ \left ( I_{bb}^{b} \right )^{-1} \left ( \frac{\partial}{%
\partial \psi_{j} } I_{bb}^{b} \right ) \right \}, \\
- \frac{\partial^{2}}{\partial \psi_{j} \partial \psi_{k}} \hat{\ell}_{m} (\psi)
&=& I_{\psi_{j} \psi_{k}}^{b} - I_{\psi_{j} b}^{b} \left ( I_{bb}^{b} \right
)^{-1} I_{b \psi_{k}}^{b} \\
&& \hspace{-3mm} + \frac{1}{2} \OP{tr} \bigg \{ \left ( I_{bb}^{b} \right
)^{-1} \left ( \frac{\partial^{2}}{\partial \psi_{j} \partial \psi_{k}}
I_{bb}^{b} \right ) - \left ( I_{bb}^{b} \right )^{-1} \left ( \frac{\partial%
}{\partial \psi_{j}} I_{bb}^{b} \right ) \left ( I_{bb}^{b} \right )^{-1}
\left ( \frac{\partial}{\partial \psi_{k}} I_{bb}^{b} \right ) \bigg \},
\end{eqnarray*}
for $1 \leq j, ~ k \leq p$, where $I_{xy}^{b} = - \frac{\partial^{2}}{%
\partial x \partial y^{\OP{T}}} \ell_{e} (\psi, b) \big |_{b = \tilde{b}}$. On the other hand, with $\ell_{e} = \ell_{e}(\psi, w)$, 
\begin{eqnarray*}
\frac{\partial}{\partial \psi} \ell_{e} ( \psi, \tilde{w} ) &=& \frac{%
\partial \ell_{e}}{\partial \psi} \Big |_{w = \tilde{w}}, \\
\left \{ -\frac{\partial^{2}}{\partial \psi \partial \psi^{\OP{T}}}
\ell_{e} ( \psi, \tilde{w} ) \right \}^{-1} &=& I_{e}^{\psi \psi},
\end{eqnarray*}
where 
\[
\begin{pmatrix}
I_{e}^{\psi \psi} & I_{e}^{\psi w} \\ 
I_{e}^{w \psi} & I_{e}^{w w }
\end{pmatrix}
= 
\begin{pmatrix}
I_{e, \psi \psi} & I_{e, \psi w} \\ 
I_{e, w \psi} & I_{e, w w }
\end{pmatrix}
^{-1}, ~ 
\begin{pmatrix}
I_{e, \psi \psi} & I_{e, \psi w} \\ 
I_{e, w \psi} & I_{e, w w }
\end{pmatrix}
= 
\begin{pmatrix}
- \frac{\partial^{2} \ell_{e} }{\partial \psi \partial \psi^{\OP{T}}} & - 
\frac{\partial^{2} \ell_{e} }{\partial \psi \partial w^{\OP{T}}} \\ 
- \frac{\partial^{2} \ell_{e} }{\partial w \partial \psi^{\OP{T}}} & - 
\frac{\partial^{2} \ell_{e} }{\partial w \partial w^{\OP{T}}}
\end{pmatrix}
_{w = \tilde{w}}. 
\]

\end{document}